\documentclass[12pt,psfig,a4]{article}
\usepackage{geometry}
\usepackage{graphics,algorithmicx,algpseudocode,algorithm,pseudocode}
\usepackage{graphicx,caption,subcaption}
\usepackage[bookmarks=false]{hyperref}
\usepackage{setspace}
\usepackage{amsmath,amssymb,amsthm,bbding}
\newcommand{\Lyx}{L\kern-.1667em\lower.25em\hbox{y}\kern-.125emX\spacefactor1000}

\newcommand{\sg}{\texttt{SortedGreedy}}
\newcommand{\gr}{\texttt{Greedy}}
\def\easytfoursymbol#1{\mathord{\mathchoice
  {\mbox{\fontsize\tf@size\z@\usefont{T4}{\rmdefault}{m}{it}\char#1}}
  {\mbox{\fontsize\tf@size\z@\usefont{T4}{\rmdefault}{m}{it}\char#1}}
  {\mbox{\fontsize\sf@size\z@\usefont{T4}{\rmdefault}{m}{it}\char#1}}
  {\mbox{\fontsize\ssf@size\z@\usefont{T4}{\rmdefault}{m}{it}\char#1}}
}}
\newtheorem{mytheorem}{Theorem}
\newtheorem{mylemma}{Lemma}
\newcommand{\BigO}[1]{\ensuremath{\operatorname{O}\bigl(#1\bigr)}}

\makeatother

\theoremstyle{plain}
\begin{document}
\bibliographystyle{unsrt} 
\pagestyle{plain} 
\pagenumbering{arabic}
%\rmfamily

\title{Balancing indivisible real-valued loads in arbitrary networks}
\author{ \"Omer Demirel, Ivo F.~Sbalzarini  \\
  \multicolumn{1}{p{.7\textwidth}}{\centering\emph{MOSAIC Group, Center for Systems Biology at Dresden, Max Planck Institute of Molecular Cell Biology and Genetics, Pfotenhauerstr.~108, D--01307 Dresden, Germany.}}}
\maketitle

% Article starts here

\begin{abstract} 
In parallel computing, a problem is divided into a set of smaller tasks that are distributed across multiple processing elements. Balancing the load of the processing elements is key to achieving good performance and scalability. If the computational costs of the individual tasks vary over time in an unpredictable way, dynamic load balancing aims at migrating them between processing elements so as to maintain load balance. During dynamic load balancing, the tasks amount to indivisible work packets with a real-valued cost. For this case of indivisible, real-valued loads, we analyze the balancing circuit model, a local dynamic load-balancing scheme that does not require global communication. We extend previous analyses to the present case and provide a probabilistic bound for the achievable load balance. Based on an analogy with the offline balls-into-bins problem, we further propose a novel algorithm for dynamic balancing of indivisible, real-valued loads. We benchmark the proposed algorithm in numerical experiments and compare it with the classical greedy algorithm, both in terms of solution quality and communication cost. We find that the increased communication cost of the proposed algorithm is compensated by a higher solution quality, leading on average to about an order of magnitude gain in overall performance.
%\sg{}\cite{Demirel:2013} is a sorting-based greedy algorithm that is designed originally to solve weighted balls-into-bins problem in an offline setting. In this manuscript we use \sg{} to dynamically balance indivisible, fixed-weight loads in random connected networks. Although \sg{} is originally designed for solving balls-into-bins problem, it produces a much finer load balance than a simple yet powerful \gr{} method in our numerical tests. First, using similar ideas presented by Sauerwald and Sun \cite{sauerwald2012tight} we show that the same bounds derived for DLB algorithms that balance indivisible, tokens can be derived for DLB methods for balancing constant real-valued loads in randomly connected network topologies.  In addition to that we compare the performance of \sg{} against \gr{} method. From our numerical results we conclude that in a setting where every load is free to move to a neighboring node \sg{} can reduce an initial load discrepancy in a randomly connected network by up to 1605x while providing a 371.31x better result than \gr{}. In a more realistic setting where some loads are not allowed to move freely, \sg{} outperforms \gr{} by up to 45x. By also taking the required load movements between neighboring nodes into account we show that the relative performance of \sg{} is on average more than 20x better than \gr{}.

\end{abstract} 


\section{Introduction}
Dynamic load balancing (DLB) aims at evenly distributing loads (e.g., computational tasks) among processing elements (e.g., computer cluster nodes) that are connected by communication links of some topology. The goal is that the time to completion on all processors be equal, hence minimizing the total execution time of the complete set of tasks. DLB is required in situations where the time required to complete a task may change in an unpredictable way during task processing. This may render an initially well-balanced load distribution unbalanced over time. DLB aims at maintaining good load balance by dynamically (i.e., during execution time) transferring tasks from overloaded processors to underutilized ones. 

DLB can be viewed and modeled in two ways: the \textit{task view} or the \textit{processor view}. In the \textit{task view} one models the tasks and their mutual dependencies by a graph, where tasks are represented by vertices and dependencies by edges. The DLB problem then becomes an NP-complete graph-partitioning problem~\cite{gary1979computers}. Several efficient heuristics and approximation algorithms have been developed to address this problem~\cite{sanders2011engineering,karypis1999parallel,walshaw2000parallel,pellegrini2012pt}, as implemented in software libraries such as \texttt{Metis}~\cite{karypis1995metis}, \texttt{Chaco}~\cite{hendrickson1995multi}, \texttt{Jostle}~\cite{walshaw2007jostle}, and \texttt{Scotch}~\cite{pellegrini1996scotch}. Recent work in solving DLB problems using the \textit{task view} is concerned with designing multi-threaded graph partitioners~\cite{lasalle2013multi}. 

The \textit{processor view} represents the interprocessor communication network as a graph, where each vertex represents a processor and an edge is drawn between processors that are connected by a direct line of communication. The total ``weight'' of a vertex is the sum of all loads assigned to that processor. The DLB problem then corresponds to moving the loads across the graph to evenly distribute them among the vertices. Scalable DLB algorithms only move loads between adjacent processors to avoid global communication. Such are not severely affected by growing network size. The quality of a load distribution is measured by the \textit{discrepancy}, which is the difference between the heaviest and lightest node in the network.

Within the \textit{processor view}, there are two subclasses of scalable DLB algorithms that differ in the local communication model (one-to-one vs.~one-to-all neighbors): In \textit{diffusion}-based algorithms~\cite{cybenko1989dynamic,boillat1990load} the total load on a node is balanced concurrently with all nearest neighbors. Alternative models include \textit{dimension exchange}~\cite{cybenko1989dynamic} and the \textit{matching model}~\cite{ghosh1996dynamic}, where a node selects a single neighbor in each round, and they balance their loads in isolated pairs.  

Theoretical analysis of balancing arbitrarily divisible loads (i.e., the \textit{continuous} case) using diffusion-based DLB schemes has been done using spectral analysis of Markov processes on graphs~\cite{rabani1998local,roberts1997weak}. The analysis has also been extended to distributed gossip algorithms that reach a consensus~\cite{aysal2009broadcast} or perform a collective operation (e.g., averaging) in the network~\cite{boyd2006randomized}. Similar analysis applies also to a model where the loads are indivisible, unit-sized tokens (i.e., the \textit{discrete} case)~\cite{ghosh1996dynamic,muthukrishnan1998first}. Some authors used randomized rounding to quantify the difference between the continuous and the discrete case~\cite{rabani1998local,mitzenmacher2001power,friedrich2009near}. Sauerwald and Sun~\cite{sauerwald2012tight} showed tight bounds on the deviation between the continuous and the discrete case. 

The model where loads are indivisible, but real-valued has not been studied in much detail. Here, we focus on DLB of \textit{indivisible, real-valued} loads in arbitrary networks. We hence consider tasks that are indivisible and carry a constant, but not necessarily unit-sized computational cost. This realistically models atomic tasks in parallel computer systems that cannot be subdivided into smaller work packages, such as in a scientific simulation where the initial computational domain is decomposed into smaller subdomains whose sizes are fixed. At any given time, the cost of each subdomain is a real number, and subdomains cannot be subdivided by a DLB protocol. We analyze the expected discrepancy between the present case and the continuous case using techniques developed earlier~\cite{sauerwald2012tight}. With some modifications to the original theoretical framework, we develop  tight bounds for the discrepancy of balancing indivisible, real-valued loads. 

Furthermore, we propose the sorting-based algorithm \sg{} for efficiently balancing indivisible, real-valued loads between two nodes. \sg{} balances the local loads faster than a classical \gr{} method, requiring fewer iterations of the DLB protocol. In addition, \sg{} achieves better load balance.

The structure of the paper is as follows: Section~\ref{sec:matching} introduces the notation and the DLB model. In section~\ref{sec:bcm} we recapitulate the theoretical framework used to bound the expected discrepancy and present tight bounds for balancing fixed, real-valued loads. In section~\ref{sec:alg}, we introduce the \sg{} algorithm and argue why it suits the present theoretical framework. In section~\ref{sec:dlb}, we show how \sg{} and \gr{} are used in a DLB protocol. Section~\ref{sec:sims} presents numerical experiments and benchmarks. We discuss the results in section~\ref{sec:discussion} and conclude in section~\ref{sec:conclusion}.

\section{Model and Notation}
\label{sec:matching}
We do not consider \textit{diffusion}-based model and instead focus on a \textit{matching model}, namely the \textit{balancing circuit model} (BCM), which has been shown to produce better local load balance in many applications~\cite{xu1995nearest}. 

Let $G=(V,E)$ be an undirected and connected graph consisting of $n$ vertices $V$. Following established notation~\cite{rabani1998local,sauerwald2012tight} we denote an edge $[u : v]$, where $\{u,v\}\in E$ with $u<v$.  Each \textit{matching} in round $t$ of a BCM is represented by an $n\times n$ matrix $\mathbf{M}^{(t)} \succeq 0$. Moreover, if $[u : v] \in \mathbf{M}^{(t)}$,  then $\mathbf{M}_{u,u}^{(t)}=\mathbf{M}_{v,v}^{(t)}=\mathbf{M}_{u,v}^{(t)}=\mathbf{M}_{v,u}^{(t)}=1/2$. If $u$ is not matched in round $t$, then $\mathbf{M}_{u,u}^{(t)}=1$ and $\mathbf{M}_{u,v}^{(t)}=0$ for all $v\neq u$. In BCM, two matched nodes $u$ and $v$ try to balance their loads as evenly as possible in round $t$. $L$ is the total number of tasks or loads.\\

\subsection{The balancing circuit model}
In BCM, a pre-determined sequence of $d$ matchings $\mathbf{M}^{(1)}$,\, \ldots,\,$\mathbf{M}^{(d)}$ is sequentially applied such that all edges in the graph are visited at least once. The resulting \textit{round matrix} $\mathbf{M}$ \cite{rabani1998local} is defined as $\mathbf{M}:= \prod_{s=1}^d \mathbf{M}^{(s)}$. The $n$ eigenvalues of $\mathbf{M}$ are defined as $\lambda_1(\mathbf{M})\geq ... \geq \lambda_n(\mathbf{M})$. Moreover, $\lambda (\mathbf{M}):=\max\{|\lambda_2(\mathbf{M})|,|\lambda_n(\mathbf{M})|\}$. 
We denote the product of a sequence of matching matrices between rounds $t_1$ and $t_2$ by $\mathbf{M}^{[t_1,t_2]}:=\prod_{s=t_1}^{t_2}\mathbf{M}^{(s)}$, where $t_2\geq t_1$. We also require the Markov chain with transition matrix $\mathbf{M}$ to be ergodic, i.e., $\lambda(\mathbf{M})<1$. 
Matching matrix sequences that satisfy this condition can be obtained by an edge coloring algorithm~\cite{brelaz1979new,panconesi1997randomized}. 
The results we show here for BCM can be extended to the \textit{random matching model}, where the matching matrices are realizations of a stochastic process.

\section{Bounds for balancing different types of loads}
\label{sec:bcm}
We use the theoretical framework introduced by Sauerwald and Sun~\cite{sauerwald2012tight} to show the deviation between the continuous and indivisible real-valued weights case. In the continuous case, where loads are arbitrarily divisible, the number of rounds needed by a BCM to balance the load in an arbitrary graph with a discrepancy of $\epsilon$ is less than or equal to $\frac{4 d}{1-\lambda(\mathbf{M})} \log \left ( \frac{Kn}{\epsilon} \right)$, where $K$ is the discrepancy in the initial load assignment and $d$ is the number of matchings $\mathbf{M}^{(1)}$,\,\ldots ,\,$\mathbf{M}^{(d)}$ (\cite{rabani1998local}, Theorem 1; \cite{sauerwald2012tight}, Theorem 2.2).
If the loads are indivisible, unit-sized tokens, the discrepancy cannot be made arbitrarily small. Using BCM on an arbitrary graph, a discrepancy of $\sqrt{12\log{n}}+1$ is reached after $\BigO{d \cdot \frac{\log{(Kn)}}{1-\lambda(\mathbf{M})}}$ rounds with probability at least $1-2n^{-2}$ (\cite{sauerwald2012tight}, Theorem 2.14).

In the present case, each load is defined by a constant real number. Loads cannot be modified or subdivided, but only moved from one processor to another. 
We show the relation between the present case and the continuous case by using a slightly modified version of the theorems from Ref.~\cite{sauerwald2012tight}.

In order for the analysis to be valid, all of the following has to be satisfied:
\begin{enumerate}
	\item The maximum load is non-increasing and the minimum load is non-decreasing.
	\item The load difference between two nodes is minimized as much as possible in each matching. 
	\item The expected error on each matching edge $[u : v] \in \mathbf{M}^{(t)}$ is zero. 
	\item Lemma 2.12 of Ref.~\cite{sauerwald2012tight} holds:
	\begin{mylemma}[\cite{sauerwald2012tight}, Lemma 2.12] 
	\label{lemma:first}
		Fix two rounds $t_1<t_2$ and the load vector $x^{(t_1)}$ at the end of round $t_1$. For any family of non-negative numbers $g_{u,v}^{(s)}$ $( [u : v] \in \mathbf{M}^{(s)},$ $t_1+1\leq s \leq t_2)$, define the random variable $Z$ by $Z:=\sum_{s=t_1+1}^{t_2}\sum_{[u : v] \in \mathbf{M}^{(s)}}g_{u,v}^{(s)}  \cdot e_{u,v}^{(s)}$. Then, $\mathbb{E}[Z]=0$ and for any $\delta>0$ it holds that
		\begin{equation}
			\Pr\left [ \left | Z - \mathbb{E}[Z] \right | \geq \delta  \right  ] \leq 2 \cdot \exp \left ( \frac{\delta^{2}}{2 \sum_{s=t_1+1}^{t_2}\sum_{[u : v] \in \mathbf{M}^{(s)}} \left (g_{u,v}^{(s)} \right )^{2}}\right ).
		\end{equation}
	\end{mylemma}
	This lemma requires that the condition (3) holds: $\mathbb{E}\left [e_{u,v}^{(s)} \right]=0, \forall \{u, v\} \in \mathbf{M}^{(s)}$. This ensures $\mathbb{E}\left [ Z \right ]=0$. We are left to prove that $Z$ is concentrated around its mean by applying an appropriate concentration inequality theorem.
\end{enumerate}
Under these conditions, the upper bound on the discrepancy that can be reached by a BCM with fixed, real-valued loads is the same as the upper bound already derived for indivisible, unit-sized tokens (Theorem 2.14 in Ref.~\cite{sauerwald2012tight}):
\begin{mytheorem} (\cite{sauerwald2012tight},Theorem 2.14)
\label{theorem:result}
	Let $G$ be any graph. Then, the following statements hold:
	\begin{itemize}
		\item Let $\mathcal{M} = \left \langle  \mathbf{M}^{(1)},\mathbf{M}^{(2)},\ldots \right \rangle$ be any sequence of matchings in a BCM. If $x^{(0)}=\xi^{(0)}$, then for any round $t$ and any $\delta \geq 1$
		\begin{equation}
			\Pr\left [ \underset{w \in V}{\max}\left | x_w^{(t)} - \xi_w^{(t)}\right |   \geq \sqrt{4\delta \cdot \log{n}}\right ] \leq 2n^{-\delta + 1}.
		\end{equation}
		\item Using BCM, a discrepancy of $\sqrt{12\log{n}}+1$ is reached after $\tau_{cont}(K,1)\in \BigO{d \cdot \frac{\log{(Kn}}{1-\lambda(\mathbf{M})}}$ with probability at least $1-2n^{-2}$.
	\end{itemize}
\end{mytheorem}
The Appendix A contains a proof of this theorem in the present model and also proves that all required conditions hold for the present model and the DLB algorithms presented in the following section.

\section{Algorithms}
\label{sec:alg}
In each matching of a BCM, the goal is to distribute the loads between two nodes $u$ and $v$ as evenly as possible. A proper DLB algorithm should reduce the load imbalance in each matching. This local load balancing problem can be formalized as an \textit{offline} balls-into-bins problem~\cite{raab1998balls,berenbrink2006balanced} with two bins. The classical \textit{balls-into-bins} problem~\cite{johnson1977urn,kolchin1978random} considers the sequential placement of $m$ balls into $n$ bins such that the bins are maximally balanced. Historically, the problem is categorized by the types of balls (e.g., uniform~\cite{azar1994balanced,azar1999balanced} vs.~weighted~\cite{berenbrink2008weighted,talwar2007balanced,peres20101+,dutta2011perfectly}), by the number of bins a ball can choose from (e.g., single-choice vs.~multi-choice~ \cite{mitzenmacher2001power}), and by the number of balls (e.g., $m=n$~\cite{raab1998balls} vs.~$m>n$ or $m\gg n$~\cite{berenbrink2006balanced}). In applications such as load balancing, hashing, and occupancy problems in distributed computing~\cite{berenbrink2008weighted,berenbrink2006balanced,czumaj2003perfectly} the $d$-choice variant and its subproblem, the two-choice variant have been the main focus. In the present case of balls having individually different weights, Talwar and Wieder~\cite{talwar2007balanced} have shown that as long as the weight distribution has finite second moment, the weight difference between the heaviest and the average bin (i.e., the discrepancy) is independent of $m$. Peres \textit{et al.}~\cite{peres20101+} introduced the $(1+\beta)$-choice process analysis, and for $\beta=1$ the discrepancy has a bound $\Theta(\log \log n)$ even for the case of weighted balls. Dutta \textit{et al.}~\cite{dutta2011perfectly} introduced the $IDEA$ algorithm, which provides a constant discrepancy with high probability even in the heavily loaded case $m\gg n$. 
 
In the offline version of the problem, we are given the complete set of balls (i.e., the loads on matched nodes) \textit{a priori}. We define the discrepancy as the weight difference between the heaviest and the lightest bin. We do not restrict the distribution from which the balls sample their weights. For simplicity, we assume that a ball can be placed into any bin, thus $d=n$. We propose to initially sort the balls according to their weights and then use a greedy algorithm to place the next heaviest ball into the lightest bin. We show that even for moderate problem sizes ($m<4000$ balls) this sorting-based greedy algorithm results in a 10 to 60-fold smaller discrepancy than the na\"{i}ve \gr{} algorithm. Furthermore, using computer simulations we show that the discrepancy resulting from the sorting-based algorithm decreases exponentially with increasing $m$. The time overhead due to sorting is negligible, which makes the sorting-based algorithm also practically useful. 

%We propose using \sg{} algorithm \cite{Demirel:2013} for solving the DLB problem in arbitrary networks. It is originally designed for solving offline balls-into-bins problem which is not investigated in literature as much in detail as its online version. Assuming the loads (i.e. tasks) are not removed or added to the network, the dynamic load balancing problem is analogous to solving offline balls-into-bins problem locally and then using balancing circuit model to reach a balanced state globally. Locally, in every matching $[u : v]$ in $\mathbf{M}^{(t)}$ $u$ and $v$ try to balance their local load vectors $x_{u}^{(t)}$ and $x_{v}^{(t)}$ with each other. These load vectors consist of independent, random loads which are located on $u$ and $v$ in round $t$. We assume that the loads draw their weights from a distribution $\mathcal{D} \in [0,A], A \in \mathbb{R}^+$. \\

\subsection{\texttt{SortedGreedy}}
We are given $m$ balls with non-negative weights $W_{1\ldots m} \in \mathbb{R}^+$ that are distributed according to a (not necessarily known) probability distribution $\mathcal{D}$. After placing the $i^{\text{th}}$ ball, we denote the total weight of bin $k$ by $U_i^{(k)}$. The discrepancy after placing $i$ out of the $m$ balls then is $G_i=\max_{k}U_i^{(k)} - \min_{k} U_i^{(k)}$. The difference between consecutive discrepancies between two iterations is bounded from above by $\Delta G_{i+1} \leq W_{i+1}$ (see Appendix B for proof).

A total of $m$ balls (local loads) are sorted in order of descending weights, such that $\{l_1\geq ... \geq l_m\}$. The loads are then placed into the bins as follows: In the beginning the load with largest weight $l_1$ is placed into any of the bins with equal probability. Then, we place the next-heaviest ball into the bin with the least current sum of weights. This procedure is repeated until every ball is placed. In BCM, this balls-into-bins algorithm is applied to all edges in a distributed fashion. The pseudocode of \sg{} can be seen in algorithm \ref{code:sg}. It consists of two phases: sorting and greedy placement. \\

\begin{pseudocode}[framebox]{SortedGreedy[n]}{U_{1...n},W}
	\COMMENT{Given are a set $W$ of $m$ balls, and the bin arrays $U_{1...n}$} \\
   	\COMMENT{Sort the array in descending order (e.g. using quicksort)} \\
   	sortedW \GETS quicksort(W)  \\
	\RETURN{\CALL{Greedy[n]}{U_{1...n},sortedW}} \\
	\label{code:sg}
  \end{pseudocode}

with

\begin{pseudocode}[framebox]{Greedy[n]}{U_{1...n},W}
	\COMMENT{Given are a set $W$ of $m$ balls and the bin arrays $U_{1...n}$} \\
   	\COMMENT{Assign the first value to the first bin} \\
	U_1[1] \GETS W[1] \\
	\COMMENT{Initialize the pointers for all bins} \\
	p_{2...n} \GETS 1 \\
	\COMMENT{First bin has already one ball in it.} \\
	p_1 \GETS 2 \\
	
	\COMMENT{Give remaining $m-1$ balls sequentially to lightest bin} \\
   	\FOR i \GETS 2 \TO m \DO
		\BEGIN 
			\COMMENT{Find the ID of the lightest bin which is the one with least current sum} \\
			idx \GETS findLightestBin(U_{1...n}) \\
			U_{idx}[p_{idx}] \GETS W[i] \\
			p_{idx} \GETS p_{idx} + 1 \\			
		\END \\
	\RETURN{U_{1...n}}
	\label{code:gr}
  \end{pseudocode}

The online version of the \gr{} algorithm has previously been proposed~\cite{azar1994balanced,azar1999balanced} and extended to the weighted balls case~\cite{talwar2007balanced}. 

 If the ball weights are drawn from a uniform random distribution, we can set an upper bound on the possible lightest weight of a ball:
 \begin{mylemma}
	Consider two ball weight samplings $A$ of size $m$ and $B$ of size $m+1$. Both $A$ and $B$ draw their ball weights from the same uniform distribution $\mathcal{D} \in [0,1]$.  Then, $\Pr[\min(A)\geq \min(B)]=\frac{m}{m+1}$. Thus, $\min(A)\geq \min(B)$ with high probability. 
	\label{lemma:wn1}
\end{mylemma}
\textit{Proof.} Let $A_i$ denote a random sample in $A$. Then, $\Pr[A_i\leq\frac{1}{m}]=\frac{1}{m}$ and thus, $\Pr[\min(A)\leq\frac{1}{m}]=1$. Similarly, $\Pr[\min(B)\leq\frac{1}{m+1}]=1$. Moreover, we can say $\min(A)\leq\frac{1}{m}$ and  $\min(B)\leq\frac{1}{m+1}$ in $\mathcal{D}$. Using these, let us device another uniform distribution $\mathcal{D'} \in [0,\frac{1}{m}]$ from $\mathcal{D}$, which includes both $\min(A)$ and $\min(b)$. Then, $ \Pr[\min(A)\geq \min(B)]=\frac{\min(B)}{\min(A)}=\frac{m}{m+1}$. Taking the limit as $m$ goes to infinity yields  $\min(A)\geq \min(B)$. $\blacksquare$ \\

Using Lemma~\ref{lemma:wn1} and Eq.~\ref{eqn:mbin_switch_tightbound} we can bound $\Delta G_m$ by
 \begin{align}
   \Delta G_m \leq W_m \leq \frac{1}{m}. 
   \label{eqn:bounding2}
 \end{align}
 
Since $1\ge W_1 \geq \frac{m-1}{m} $, the final discrepancy becomes
   \begin{align}
   \nonumber
   G_m & \geq  W_1 - \sum_{i=2}^{m} W_i \\ \nonumber
   & \geq 2W_1 - \sum_{i=1}^{m} W_i \\
   & \geq 2W_1 - \sum_{i=1}^{m} \frac{1}{i}  .
   \label{eqn:gn_sg1}
 \end{align}

For $n>2$ the final discrepancy will be bigger as there will be empty intermediate bins which will receive at least one ball during the course (see Appendix B). Moreover, since our weights arrive in descending order, $W_n$ is the minimum ball weight and thus the upper bound on $\Delta G_m$ decreases with the same rate. In other words, fluctuations in $\Delta G_i$ are damped as $i\to m$. 

Using Lemma~\ref{lemma:wn1} we can state that the final discrepancy obtained by \sg{} decreases as $m$ increases. Furthermore, \sg{} always tries to minimize the discrepancy by putting the next ball into the lightest bin. These two features combined make \sg{} a robust algorithm that approximately solves offline weighted-balls-into-bins problems and gives a diminishing discrepancy as $m$ increases. 

If the weights are sampled from a uniform distribution over the interval $[0,1]$, we can use a distribution-based sorting algorithm, such as bucketsort, Proxmap-sort~\cite{standish1997data}, or flashsort~\cite{neubert1998flashsort}. Since these algorithms are not comparison-based, the $\Omega(m \log{m})$ lower bound for comparison-based sorting does not apply to them. For example, Proxmap-sort~\cite{standish1997data} has an average time complexity of $\BigO{mk}=\BigO{m}$, where $k<m$ is the content number of ``buckets" used for sorting. 
Thus, the algorithm outperforms the lower bound for comparison-based sorting for large $m$. The worst-case complexity of distribution-based sorting algorithms, however, is $\BigO{m^2}$ as $k$ approaches $m$. 
However, the probability of the worst case scenario (i.e., having $k=m$ buckets) is small since $k$ is user-defined. For flashsort, $k=0.42m$ is found a good value in empirical tests~\cite{neubert1998flashsort}.  

For non-uniform weight distributions, we resort to efficient comparison-based sorting algorithms, such as mergesort or quicksort~\cite{hoare1962quicksort}, which have an average time complexity in $\BigO{m \log{m}}$. Depending on the specific sorting algorithm, the worst-case complexity can also be in 
$\BigO{m \log{m}}$. Highly optimized implementations of these algorithms are commonly available. 

See Appendix C for simulation results and timings of \sg{}.

\section{BCM-based DLB methods}
\label{sec:dlb}
We analyze two DLB strategies based on the BCM using either \gr{} or \sg{} to balance the loads in each matching $[u : v]$. The sequence in which the algorithm visits the edges (i.e., the matchings) is given by an (in practice approximate) minimum edge-coloring algorithm. We assume that this task is done before the DLB algorithm is executed, such that the edges are colored and edges of the same color can be balanced concurrently. All edges are visited at least once. A pseudocode for this strategy is given in algorithm \ref{code:dlb}. 

\begin{pseudocode}[framebox]{DLB}{L,E_{colored},algorithm,k}

	\COMMENT{Given are global load vector $\texttt{Load}$,list of edges to visit $E_{colored}$ ...} \\
	\COMMENT{... an $algorithm$ to balance the loads on a selected edge ...}  \\
	\COMMENT{... and the number of iterations of $DLB$: $k$} \\
	\FOR i \GETS 1 \TO k \DO
	\BEGIN
		\FOR j \GETS 1 \TO length(E_{colored}) \DO
			\BEGIN
				\COMMENT{Find which vertices are connected by that edge} \\
				\{u, v\} \GETS getVertices(E_{colored}[j])\\
				
				\IF algorithm == \sg{}  \THEN
				\texttt{Load} \GETS \CALL{\sg{}}{u,v,\texttt{Load}} \\
			\ELSEIF algorithm == \gr{}  \THEN
				\texttt{Load} \GETS \CALL{\gr{}}{u,v,\texttt{Load}} \\
				
			\END \\
	\END \\

	\RETURN{}
	\label{code:dlb}
  \end{pseudocode}

\section{Simulation results}
\label{sec:sims}
\begin{figure}
\centering	
	\begin{subfigure}[b]{0.32\textwidth}
		\centering
	      	\includegraphics[width=\textwidth]{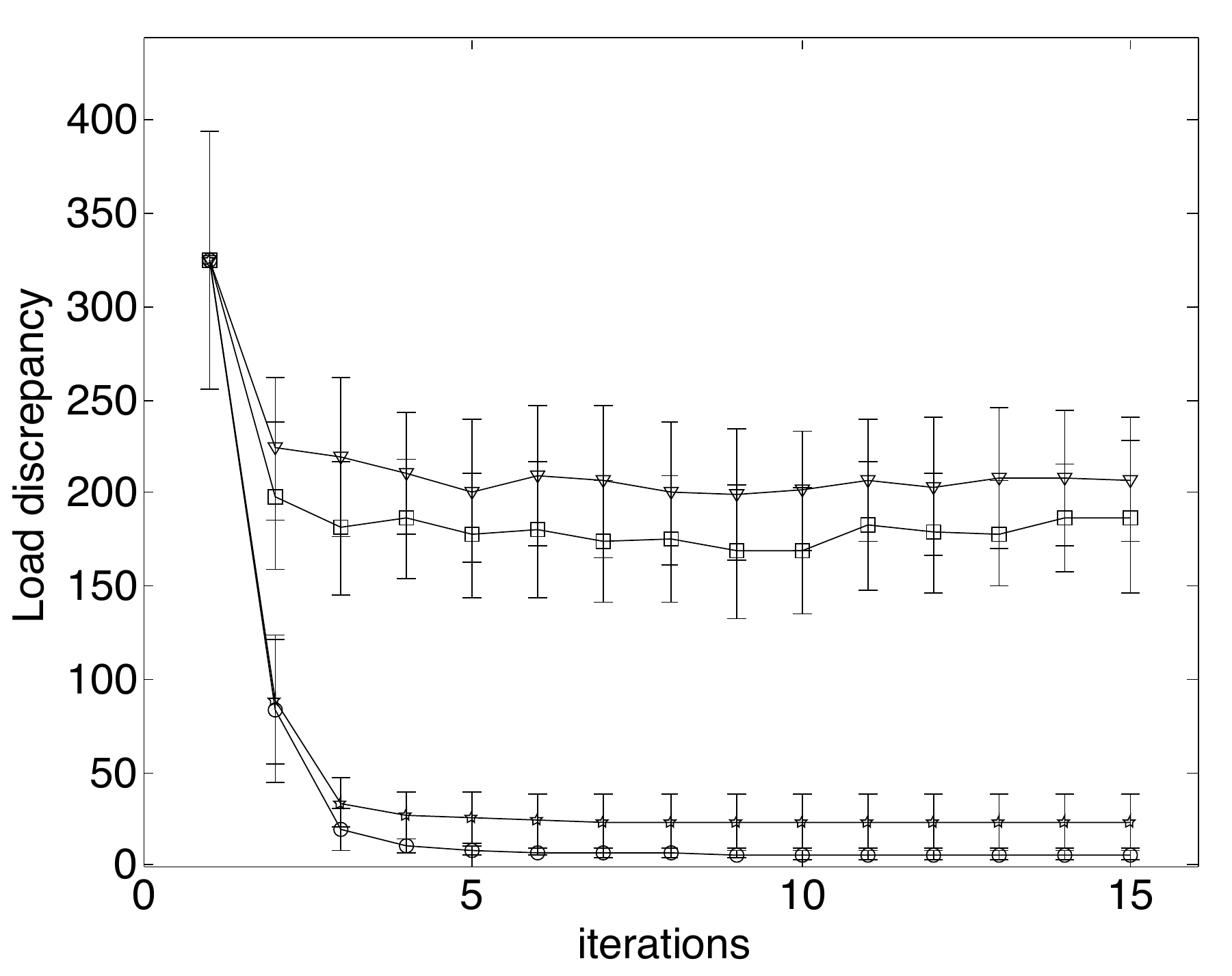}
		\caption{$n=16$, $L=160$}
		\label{fig:16_d160}
	\end{subfigure}
%	\begin{subfigure}[b]{0.32\textwidth}
%		\centering
%	      	\includegraphics[width=\textwidth]{plots/32/32_ed320}
%		\caption{$n=32$, $L=320$}
%		\label{fig:32_d320}
%	\end{subfigure}
	\begin{subfigure}[b]{0.32\textwidth}
		\centering
	      	\includegraphics[width=\textwidth]{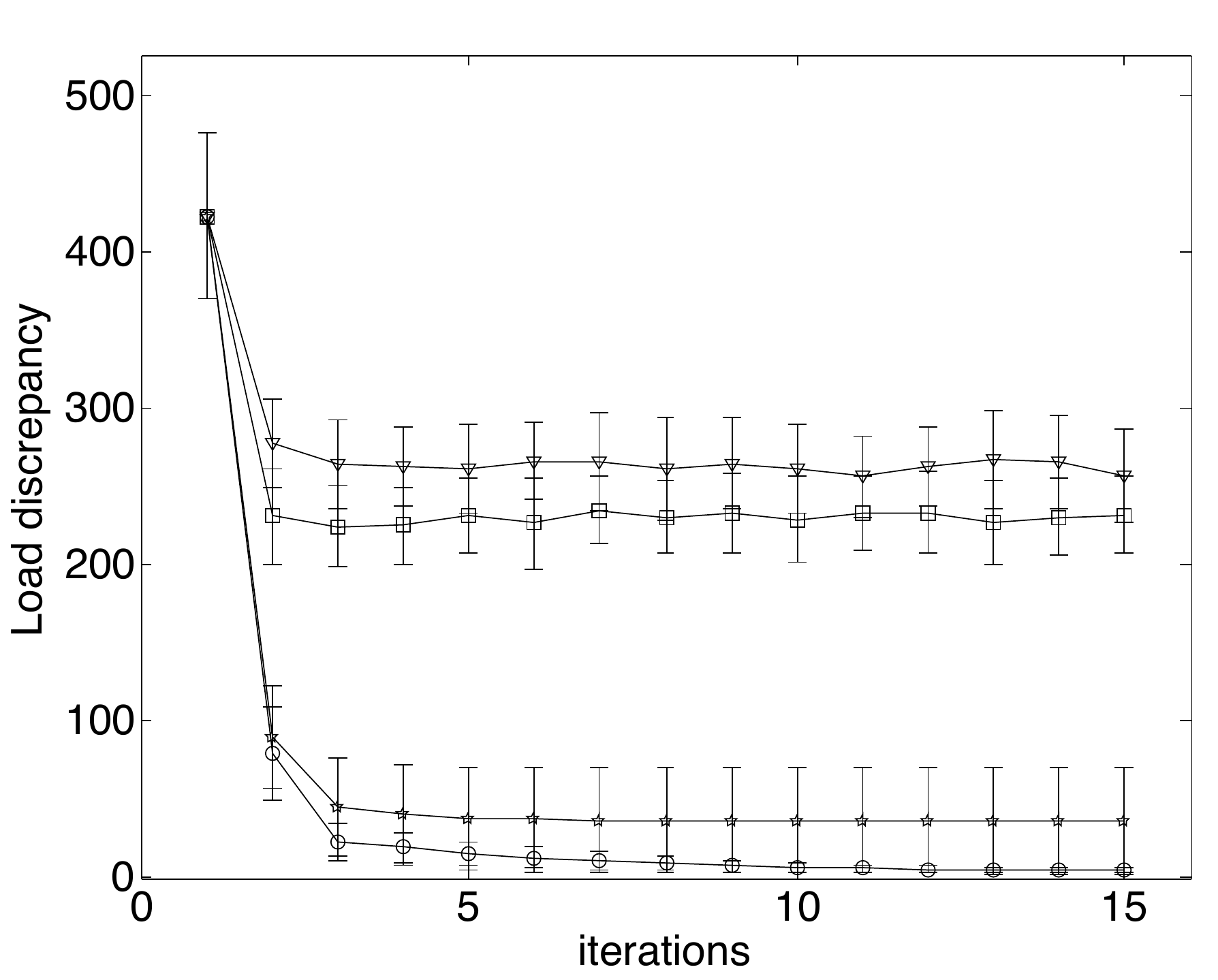}
		\caption{$n=64$, $L=640$}
		\label{fig:64_d640}
	\end{subfigure}
	\begin{subfigure}[b]{0.32\textwidth}
		\centering
	      	\includegraphics[width=\textwidth]{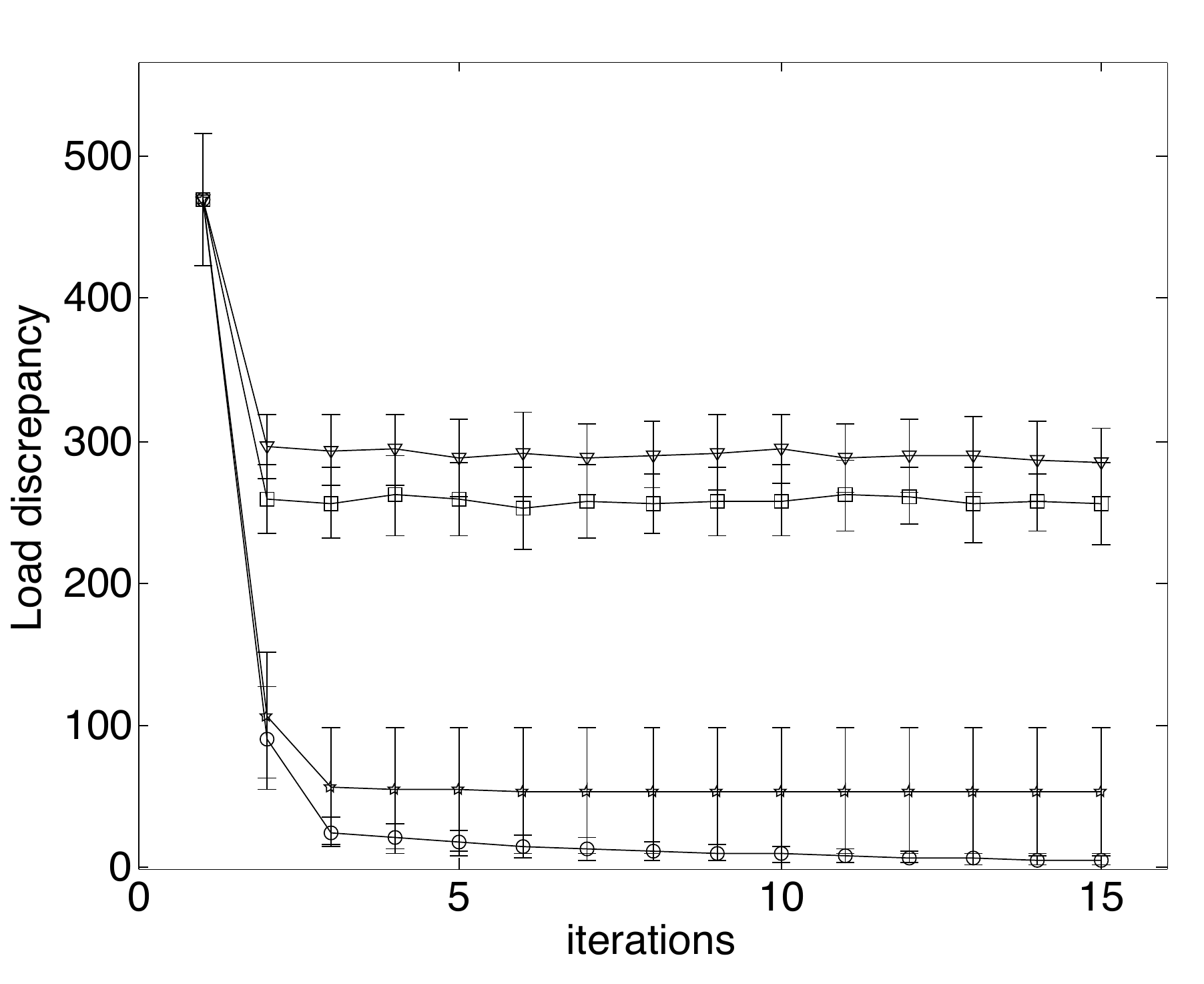}
		\caption{$n=128$, $L=1280$}
		\label{fig:128_d1280}
	\end{subfigure}
 	
	\begin{subfigure}[b]{0.32\textwidth}
		\centering
	      	\includegraphics[width=\textwidth]{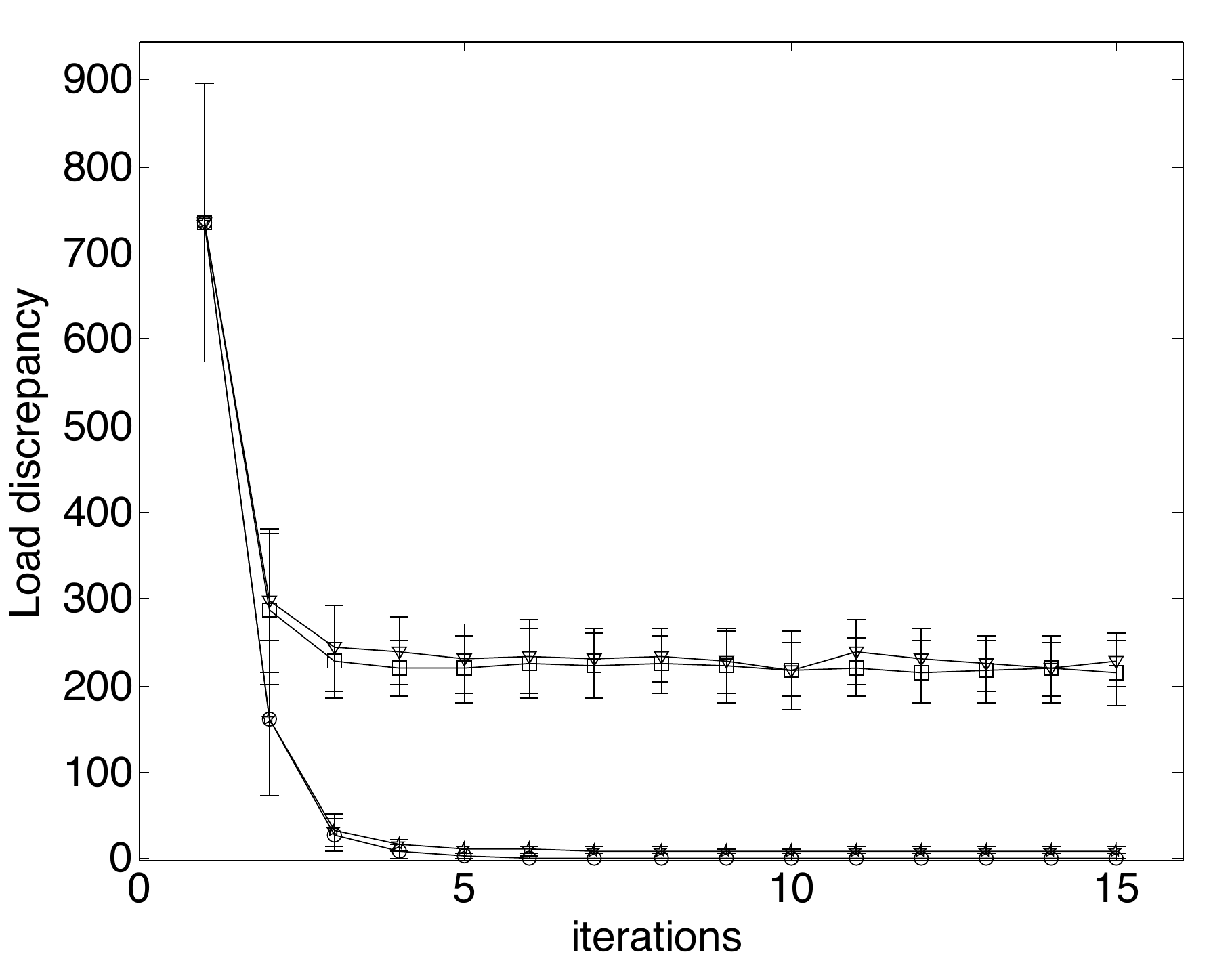}
		\caption{$n=16$, $L=800$}
		\label{fig:16_d800}
	\end{subfigure}
%	\begin{subfigure}[b]{0.32\textwidth}
%		\centering
%	      	\includegraphics[width=\textwidth]{plots/32/32_ed1600}
%		\caption{$n=32$, $L=1600$}
%		\label{fig:32_d1600}
%	\end{subfigure}
	\begin{subfigure}[b]{0.32\textwidth}
		\centering
	      	\includegraphics[width=\textwidth]{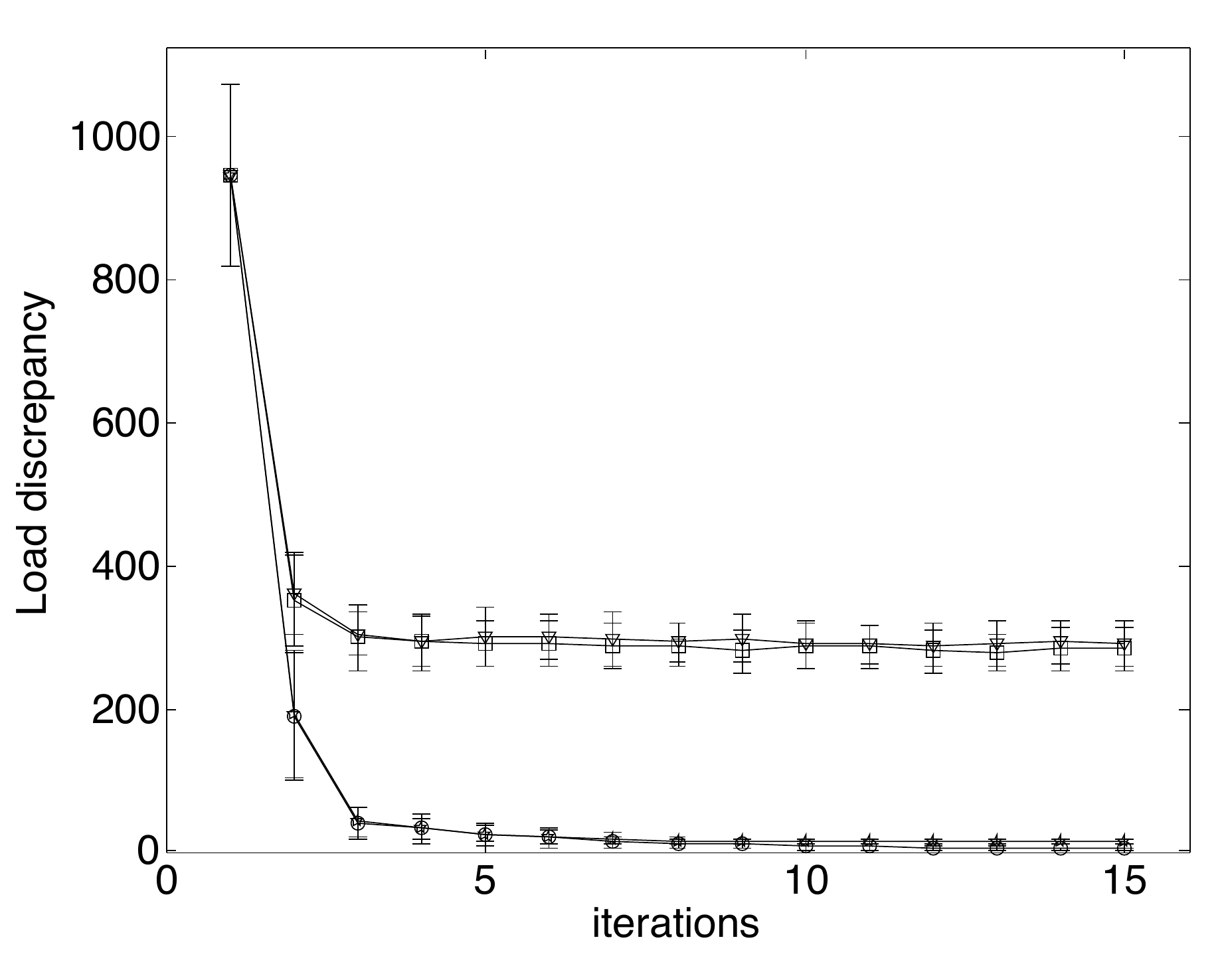}
		\caption{$n=64$, $L=3200$}
		\label{fig:64_d3200}
	\end{subfigure}
	\begin{subfigure}[b]{0.32\textwidth}
		\centering
	      	\includegraphics[width=\textwidth]{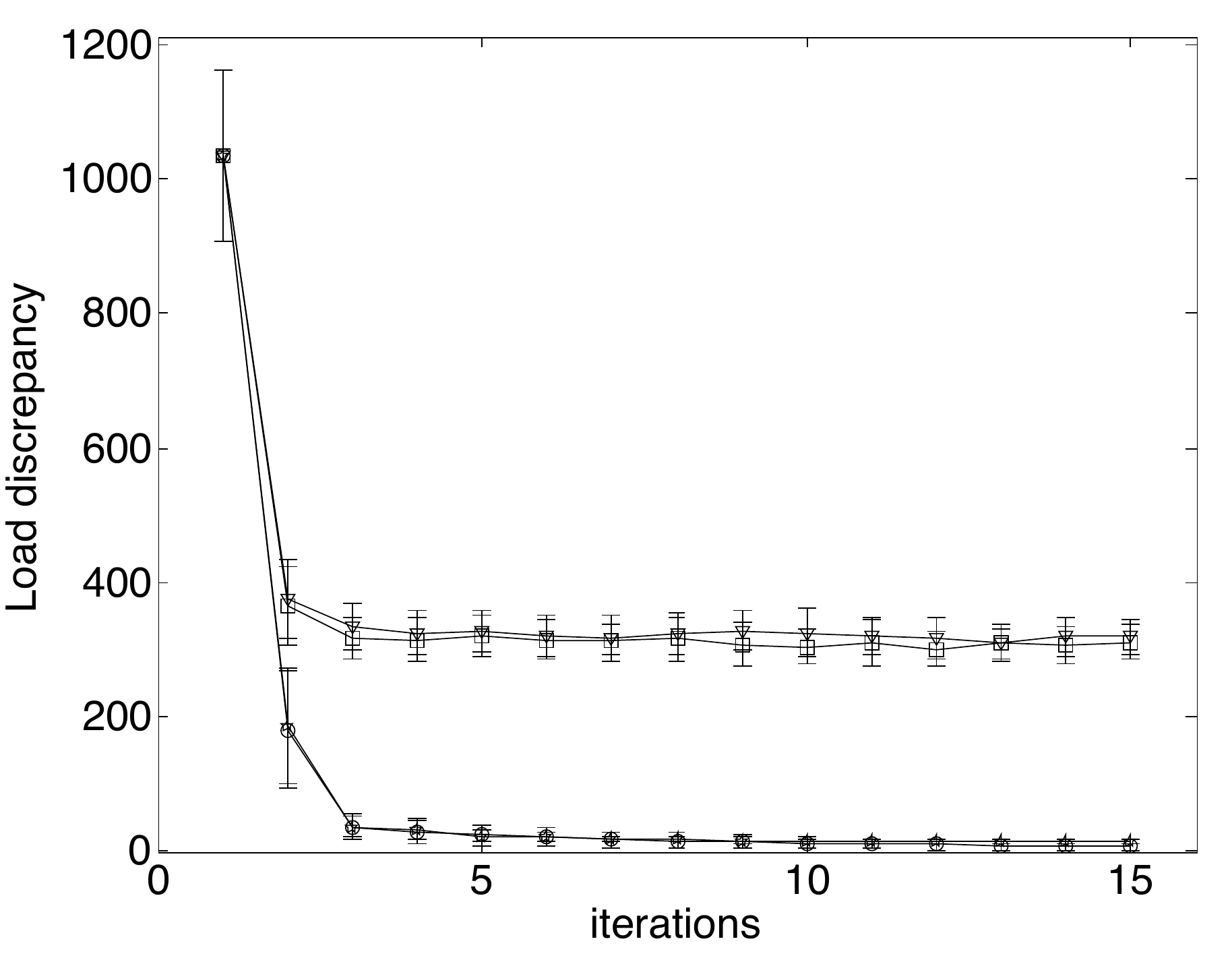}
		\caption{$n=128$, $L=6400$}
		\label{fig:128_d6400}
	\end{subfigure}
	\\
	\begin{subfigure}[b]{0.32\textwidth}
		\centering
	      	\includegraphics[width=\textwidth]{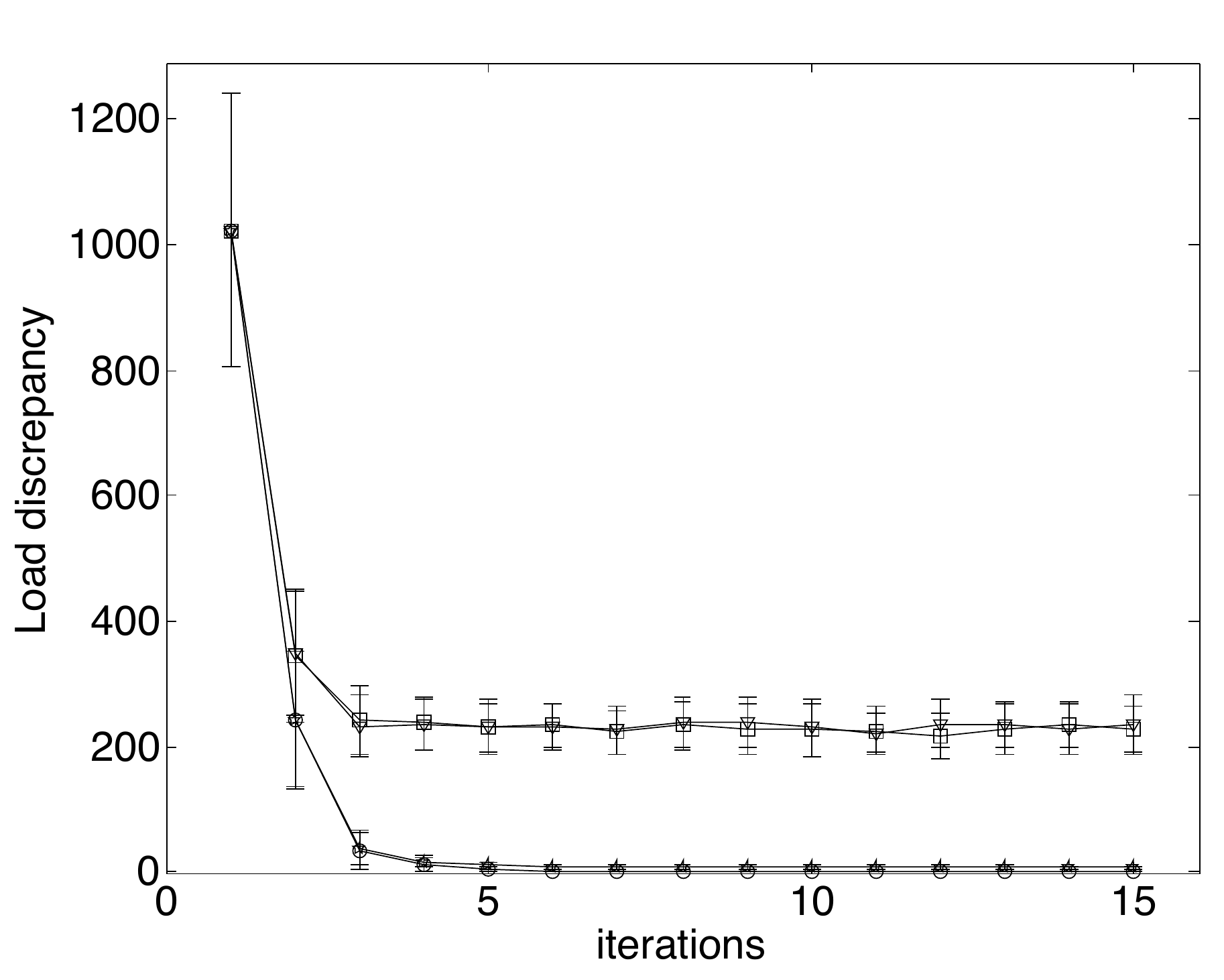}
		\caption{$n=16$, $L=1600$}
		\label{fig:16_d1600}
	\end{subfigure}
%	\begin{subfigure}[b]{0.32\textwidth}
%		\centering
%	      	\includegraphics[width=\textwidth]{plots/32/32_ed3200}
%		\caption{$n=32$, $L=3200$}
%		\label{fig:32_d3200}
%	\end{subfigure}
	\begin{subfigure}[b]{0.32\textwidth}
		\centering
	      	\includegraphics[width=\textwidth]{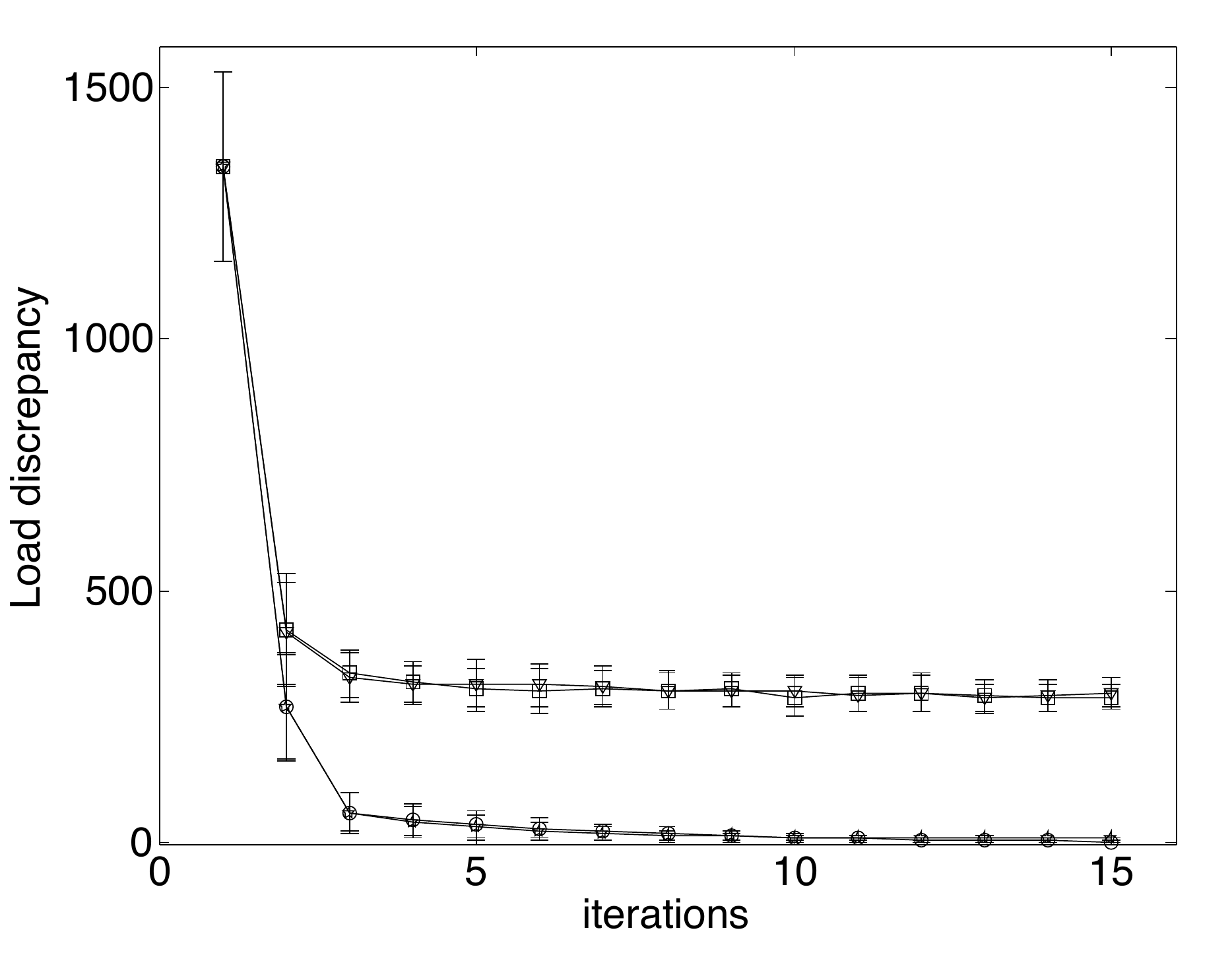}
		\caption{$n=64$, $L=6400$}
		\label{fig:64_d6400}
	\end{subfigure}
	\begin{subfigure}[b]{0.32\textwidth}
		\centering
	      	\includegraphics[width=\textwidth]{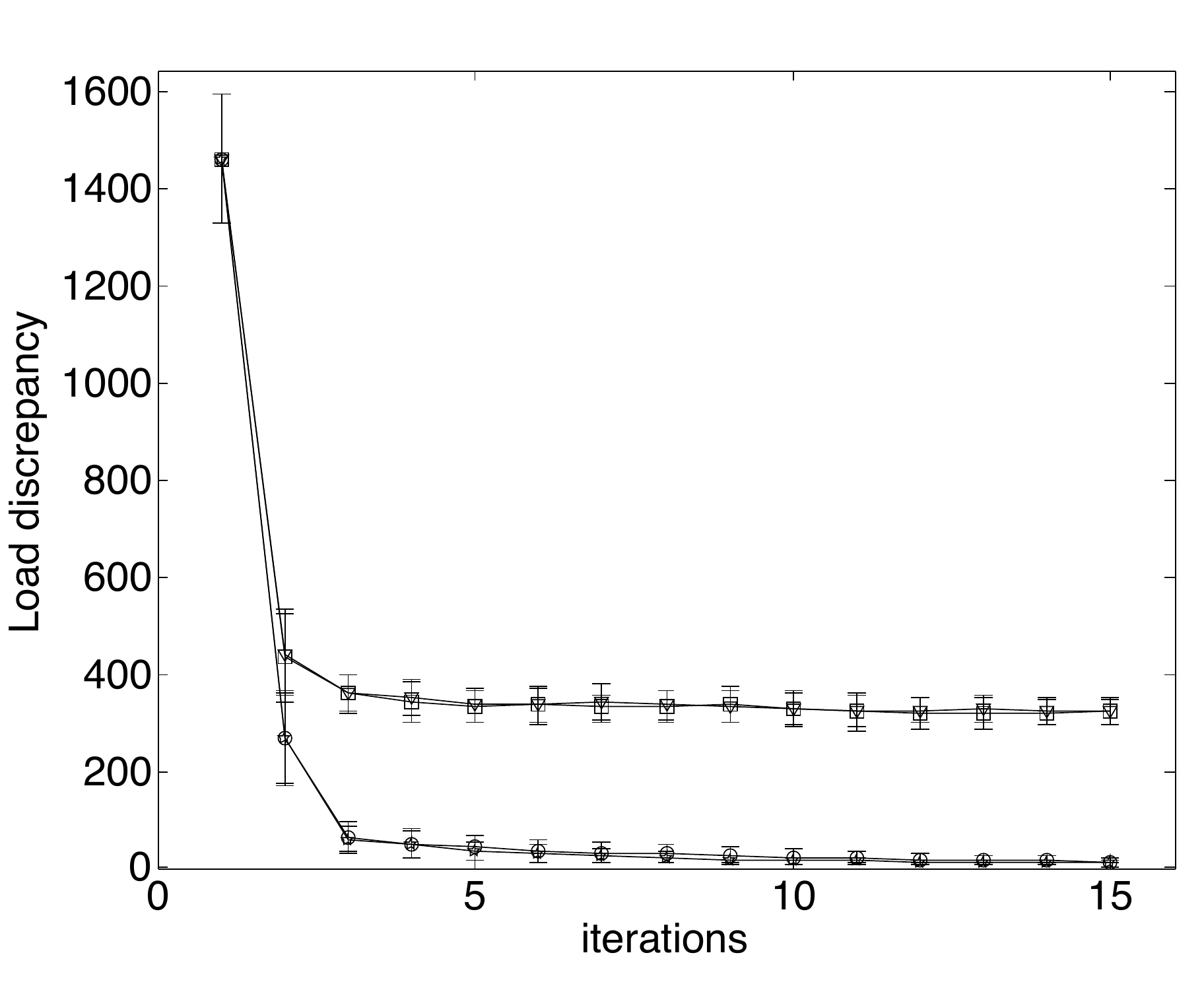}
		\caption{$n=128$, $L=12800$}
		\label{fig:128_d12800}
	\end{subfigure}
	\\

	\caption{Simulation results for dynamic balancing of indivisible, real-weight loads in randomly connected networks. Two algorithms are applied in two cases each: \sg{} with full mobility ($\circ$) and with partial mobility ($\star$), and \gr{} with full mobility ($\triangledown$) and with partial mobility  ($\square$). In (a)--(c) the number of loads is 10 times larger than the number of processors, in (d)--(f) it is 50 times larger, and in (g)--(i) it is 100 times larger.}
	\label{fig:random_con}
\end{figure}

We perform numerical experiments to illustrate the behavior of the algorithms in several scenarios. In our benchmarks, the network size $n$ ranges from 4 to 128. Edges are randomly drawn until the graph is connected. For each network size we place 10, 50, or 100 loads on each node, where loads sample their weights from an uniform random distribution over $[0,100]$. This reflects both fine-grained and course-grained domain-decomposition settings where the initial load imbalance is randomly set. We also show how increasing network size affects the present DLB algorithms. We repeat each experiment 50 times and plot the average discrepancy values along with their standard deviations in Fig.~\ref{fig:random_con}. The same graphs and initial load distributions are used for both \sg{} and \gr{}.

%\gr{} reduces the initial discrepancy up by 4-times and no significant change can be seen after the third iteration. For high $L/n$ ratios \gr{} does not seem to be influenced by the chosen mobility model (i.e. partial vs. full). 
%
%As expected, if some loads are fixed to their hosting nodes the discrepancy is usually much higher than the case where all loads are eligible to move. \sg{} performs better in all-mobile-loads tests than its semi-mobile counter part. It also performs better than \gr{} in all tested simulations where in few iterations the reduced discrepancy is up to 10 times lower than the one by \gr{} depending on the network connectivity as well as number of loads. The discrepancy ratio between  \sg{} and gr{} can go up to 80 even after five iterations and after 15 DLB iterations the discrepancy could be as big as 328.43 in case where $L=32$ and $n=100$. Interestingly, when some loads are fixed \gr{} performs slightly better than \gr{} in its all-mobile-load test case. This might be explained by the fact that the number of mobile loads for each DLB operation is decreased which could also limit the number of eligible moves for \gr{}.\\

\subsection{Mobility of loads}

While all loads are constant real numbers, it may not be practically feasible to move each load in any given BCM matching. This situation is frequently encountered in practice, for example in numerical simulations where certain biss need to stay on a given processor to maintain processor-neighborhood relationships. We denote as \textit{full mobility} the case where all loads are free to move, and as \textit{partial mobility} the case where some loads are pinned to their current processor. Assuming that there are $m$ loads on node $n_i$ we uniformly at random set $r\in[1,\,\ldots,\,l-1]$ of them to be immobile and simulate the algorithm behavior.

The full mobility case leads to lower discrepancy in all cases. We observe that \gr{} can reduce the initial discrepancy by at most 4.5-fold, which is the case for $L=12800$ and $n=128$ with full mobility. With partial mobility, the maximum discrepancy reduction observed is 4.7-fold for $L=3200$ and $n=32$. For the same configurations \sg{} reduces the discrepancy by 116-fold (with full mobility) and 132-fold (with partial mobility), respectively. Across all simulations \sg{} yields on average a 21-fold lower discrepancy than \gr{} when load mobility is restricted. With full mobility, the average discrepancy reached by \sg{} is 135-fold lower than that of \gr{}. \sg{} thus decreases the initial discrepancy on average by a factor of 1600, hence significantly improving load balance.

\subsection{Number of load movements per matching}

An important metric that is closely related to the scalability of a distributed algorithm is the communication cost. Regardless of the cost model used, the communication cost is proportional to the total number of loads moved from one processor to another one. Therefore, while the discrepancy is an important metric to measure the solution quality of a DLB algorithm, the cost at which this result is obtained plays an important role. We hence measure the average number of load movements, $\alpha$, between two neighboring nodes in a matching for both \sg{} and \gr{} with the above-mentioned mobility models for different $n$ and $L$.

\begin{figure}
\centering	
	\begin{subfigure}[b]{0.45\textwidth}	
		\centering
	      	\includegraphics[width=\textwidth]{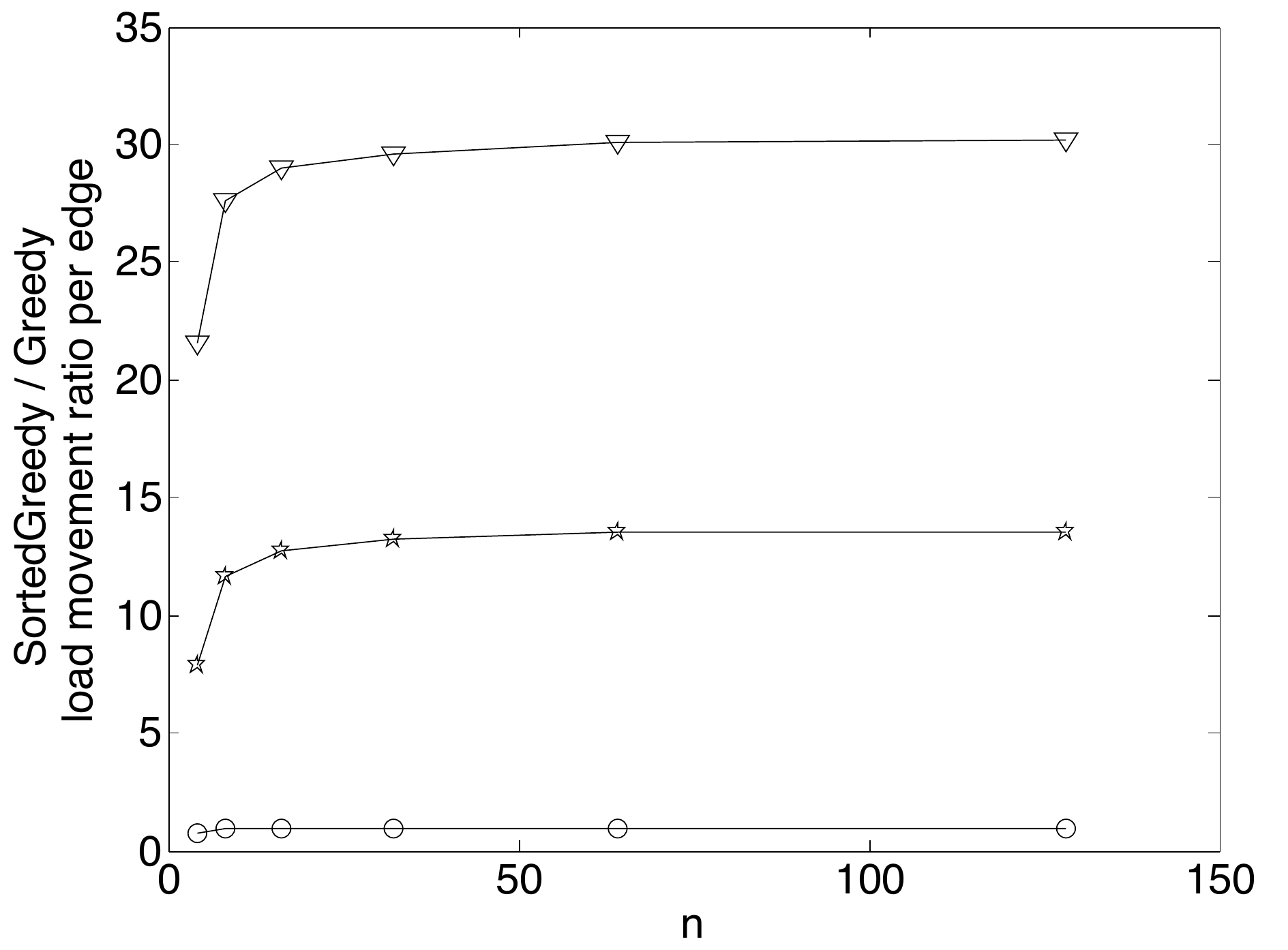}
		%\caption{}
		\label{fig:NumMovPerEdge}
	\end{subfigure}
	\begin{subfigure}[b]{0.45\textwidth}	
		\centering
	      	\includegraphics[width=\textwidth]{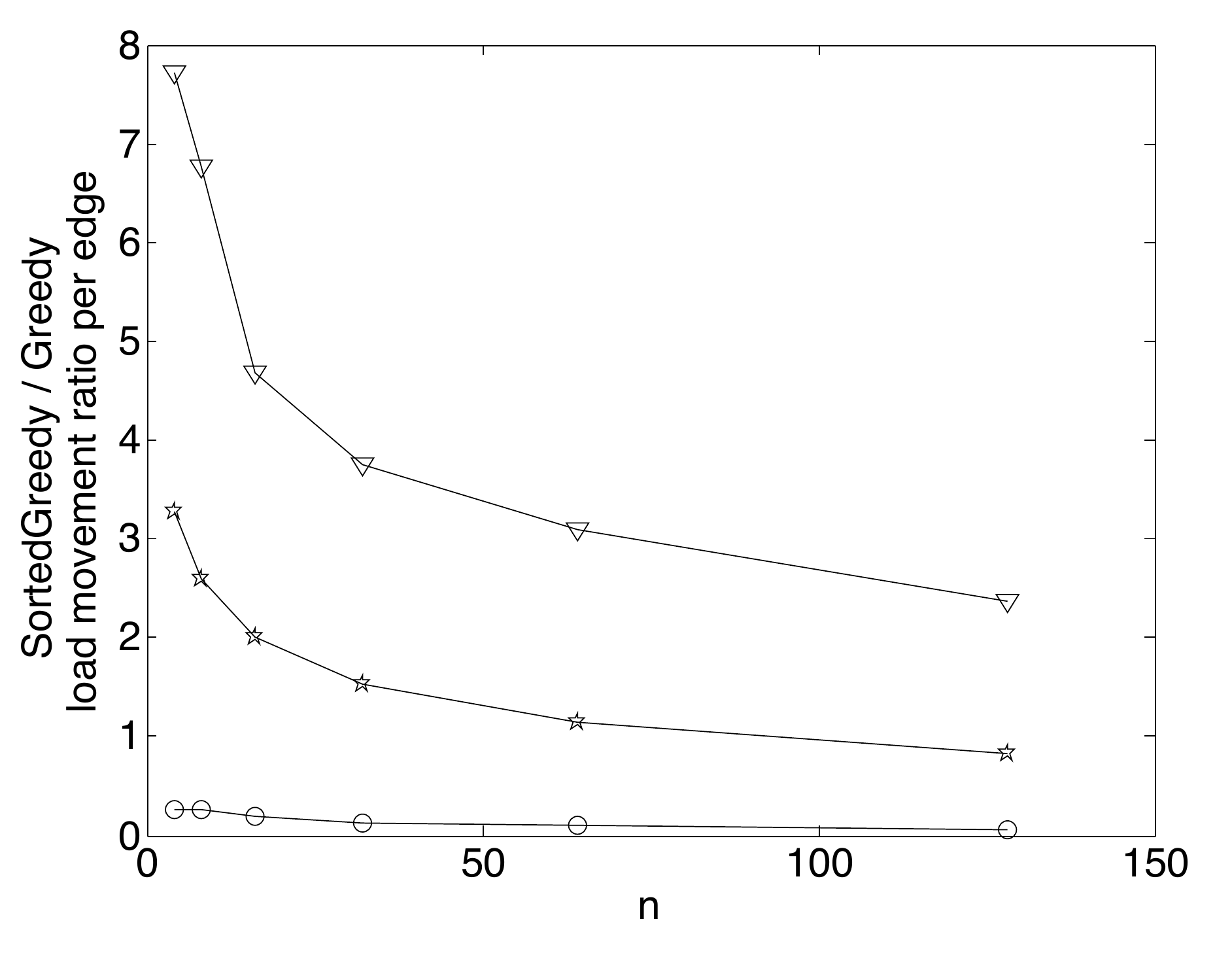}
		%\caption{}
		\label{fig:Fixed_NumMovPerEdge}
	\end{subfigure}
	\caption{The ratio of average number of load movement per edge $\frac{\sg{}}{  \gr{}}$ in the full mobility case (left) and in the partial mobility case (right) is given in different $\frac{L}{n}$ cases. Here, $\frac{L}{n}=\{10,50,100\}$ lines are shown by $\circ$, $\star$ and $\triangledown$, respectively. } 
	\label{fig:movePerEdge}
\end{figure}

As shown in Fig.~\ref{fig:movePerEdge}, \sg{} requires up to 16-fold more communication when $L/n$ is small. With full mobility and $L/n>50$, \gr{} requires up to 30 times less load movements per edge for $n=128$. The rate at which the ratio of load movements between \sg{} and \gr{} increases seems to decrease with growing network size. This indicates that there could be an approximate upper bound for the load movement ratio. In the partial mobility model, we see a decreasing load movement ratio between the two BCM variants. Even though \gr{} still requires less load movement per matching, as $L$ increases we see the load movement ratio decreasing exponentially, and for $L/n=50$ and $n=128$, \sg{} needs less load movements. On average, however, \gr{} moves 14 times (full mobility) or 2 times (partial mobility) less loads than \sg{}.
%IFS: so \sg is more expensive than \gr? Please check that all ratios are correct and sg/gr are not confused anywhere. 
%OD: I fixed the confusion. Yes, \sg is more expensive

\section{Discussion}
\label{sec:discussion}
\begin{figure}
\centering	
	\begin{subfigure}[b]{0.44\textwidth}	
		\centering
	      	\includegraphics[width=\textwidth]{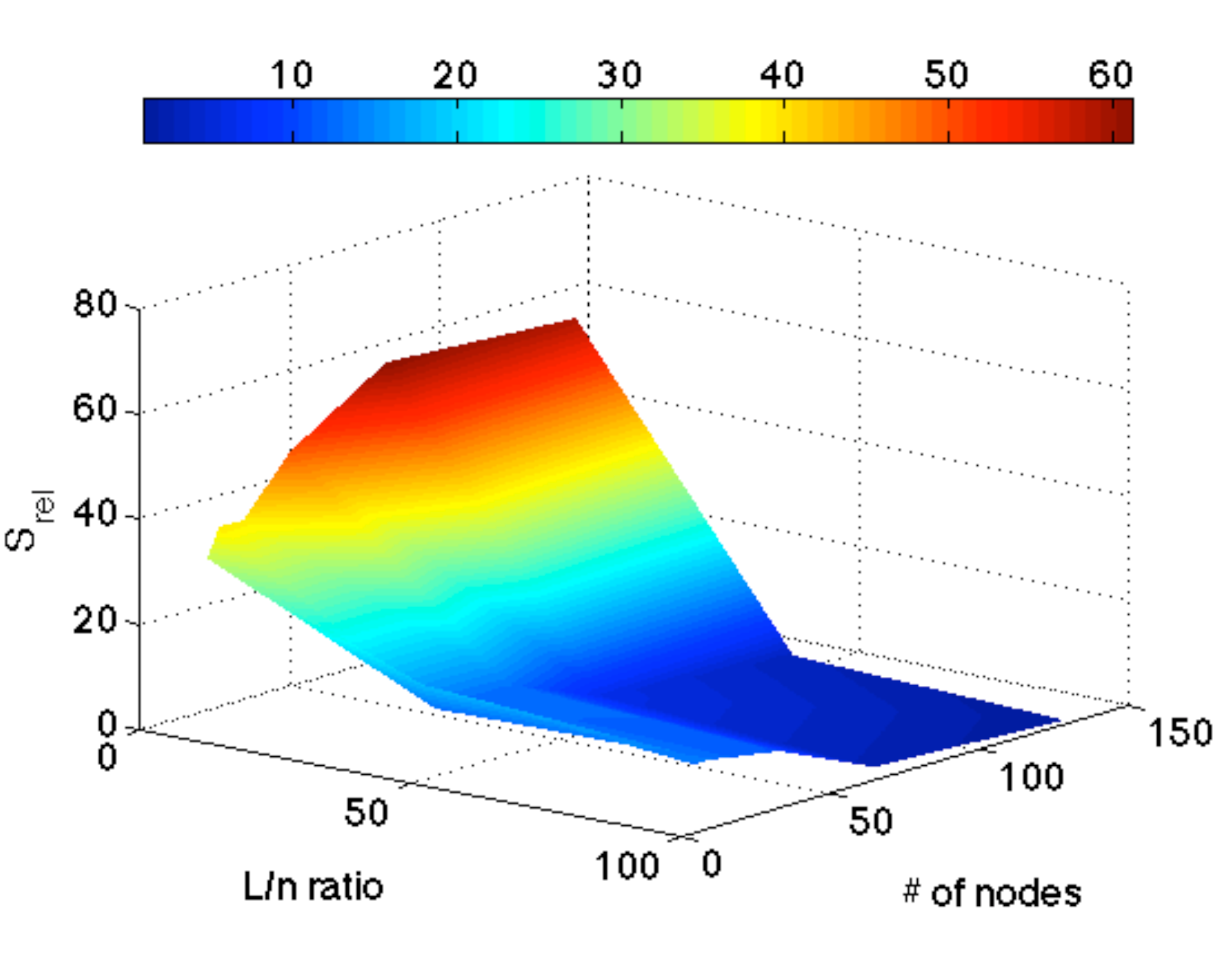}
		\caption{Full mobility}
		\label{fig:Srel1}
	\end{subfigure}
	\begin{subfigure}[b]{0.44\textwidth}	
		\centering
	      	\includegraphics[width=\textwidth]{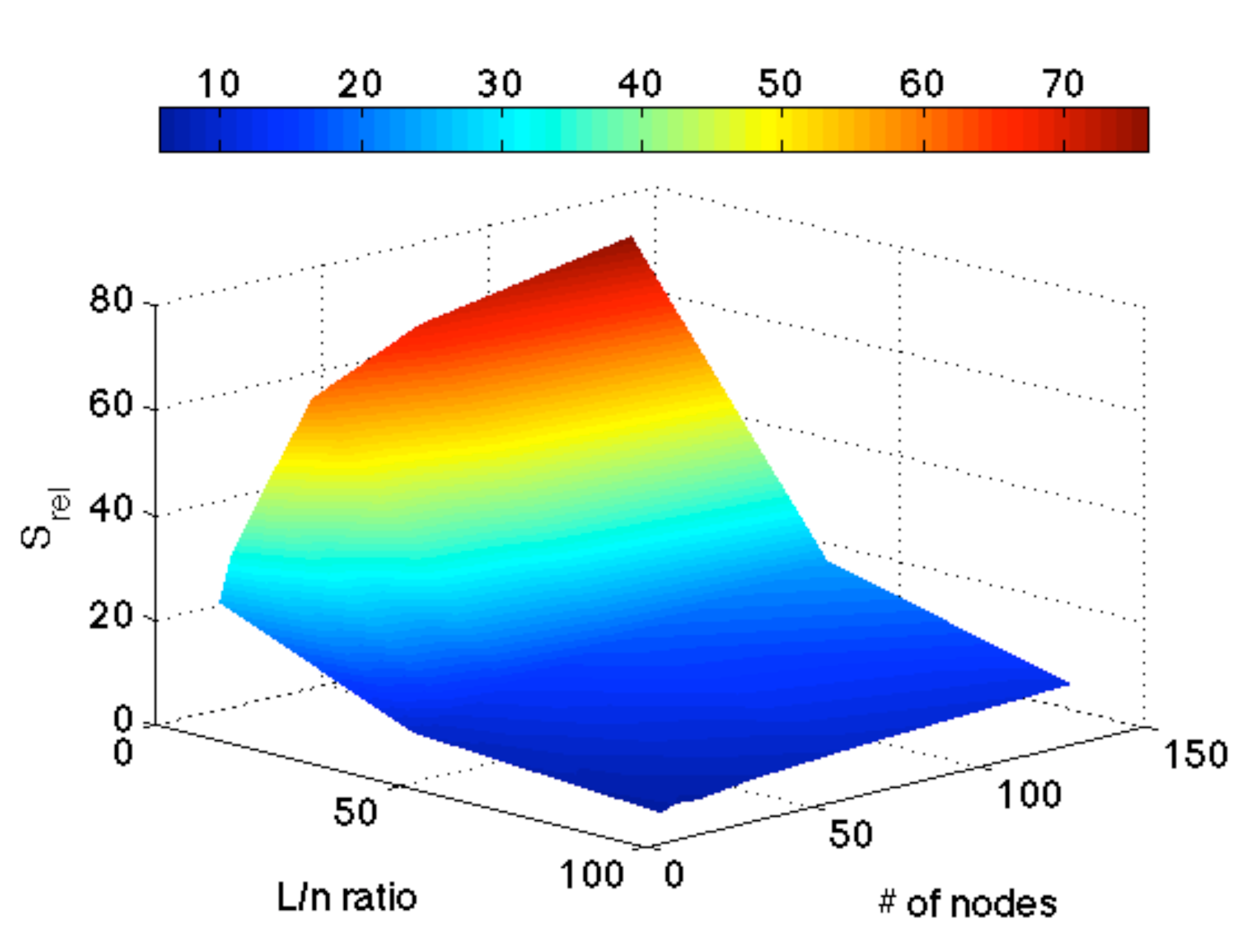}
		\caption{Partial mobility}
		\label{fig:Srel2}
	\end{subfigure}
	\caption{Relative figures of merit of \sg{} over \gr{}. \sg{} outperforms \gr{} by up to a factor of 75 and on average it is 22--24-fold better. The performance different is larger for lower $L/n$ ratios in large networks. } 
	\label{fig:Srel}
\end{figure}

Our numerical tests show that in both load mobility cases \sg{} better results than \gr{} in terms of the achieved discrepancy reduction. This comes at a cost of an on average 14-fold higher communication overhead in \sg{} than in \gr{}. For partially mobile loads, however, the communication overhead of \sg{} is only on average 2-fold larger than that of \gr{}. We formulate the following figure of merit for a BCM-based DLB algorithm:
\begin{equation}
	S = p\cdot\frac{disc}{\alpha} \, , 
\end{equation}
where $p\in \mathbb{R}^+$ is the relative importance of $disc$ over $\alpha$, $disc$ is the discrepancy reduction ratio between the initial discrepancy and final discrepancy achieved by the DLB algorithm, and $\alpha$ is the total number of load movements required to do so. The relative figure of merit $S_{\text{rel}}$ of \sg{} over \gr{} is:
\begin{equation}
	S_{rel} = \frac{S_{\text{\sg{}}}}{S_{\text{\gr{}}}} = \frac{p\cdot\frac{disc_{\text{\sg{}}}}{\alpha_{\text{\sg{}}}}}{p\cdot\frac{disc_{\text{\gr{}}}}{\alpha_{\text{\gr{}}}}}=\frac{\frac{disc_{\text{\sg{}}}}{\alpha_{\text{\sg{}}}}}{\frac{disc_{\text{\gr{}}}}{\alpha_{\text{\gr{}}}}} \, . 
\end{equation}
It is plotted for both load mobility models in Fig.~\ref{fig:Srel}. The average figure of merit of \sg{} is 22-fold or 24-fold better than that of \gr{} under full or partial load mobility, respectively. 

When the plot on the right in Fig.~\ref{fig:movePerEdge} is extrapolated, it is also to note that for bigger networks ($n>128$) with partial load mobility, \sg{} is expected to have lower load movement than \gr{}, which eliminates the only disadvantage of \sg{} against \gr{}.

\section{Conclusion and future work}
\label{sec:conclusion}
We show tight bounds on the expected discrepancy when a BCM is used to balance indivisible, real-valued loads in arbitrary networks. Our theoretical considerations closely followed prior work on the discrete case of unit-sized loads~\cite{sauerwald2012tight}. We showed that the bounds derived for the discrete case also apply in the case of real-valued loads if (i) the maximum load in the network is non-increasing and the minimum load is non-decreasing; (ii) a DLB algorithm is used that balances the local loads in each matching as much as it can; (iii) the expected error is zero on a matched edge; and (iv) the concentration bounds of the error are adjusted from the fixed-weight case.

We analyzed theoretically the offline weighted balls-into-bins problem and discussed two different approaches, namely \gr{} and \sg{}. The performance of \gr{} is not unreliable due to the sequential allocation of random $m$ balls into $n$ bins and the final resulting discrepancy $G_m$ depends on the average weight of the balls. On the other hand, by sorting the input data according to the weights \sg{} yields a final discrepancy, which is reduced by \BigO{\log{m}} for $m\gg n$. Moreover, in practice \sg{} runs almost as fast as \gr{}. This makes sorting-based algorithms favorable for solving offline weighted balls-into-bins problems.

We implemented two variants of BCM-based DLB protocols using either \sg{} or \gr{} as the core load-balancing mechanism in each matching. We analyzed these algorithms using the balls-into-bins formalism and compared their complexity and solution quality. We numerically simulated the behavior of both algorithms in randomly generated connected networks with full or partial load mobility. Our numerical tests showed that in both load mobility cases \sg{} gives favorable results where the discrepancy achieved by \sg{} is on average 135-fold or 21-fold lower than that of \gr{} for full mobility and partial mobility, respectively. On the other hand, the cost of \sg{} due to load movement is on average 14-fold larger than the cost of \gr{} for full mobility and 2-fold larger for partial mobility. In the overall quality/price ratio, $S_{rel}$, \sg{} performs on average 20-fold better than \gr{} for any load mobility model. The figure of merit of a BCM protocol largely depends on the ratio $L/n$. 
%IFS: of course the really important figure would be how much the DLB costs in relation to how much time it is going to save due to better load balance. This depends on the application and the domain decomposition and can only be shown in real benchmarks of real PPM simulations. The thing is that moving a load to another processor cost much more than communicating a ghost layer. So if the ratio between ghost-layer volume and subdomain-volume is higher than 20, maybe \sg{} does not give you any benefit at all in the end.... To be kept in mind! 

Future work will be concerned with comparing \sg{}-based BCM with other DLB algorithms and extend the tests to larger network sizes. 
%IFS: good try! I assume the reviewers are anyway going to ask for this, but let's see....
In the presented simulations, we focused on the load balancing methods and theory in an ideal setting. We neglected the specifics of the computer system and the parallel application in order to highlight some general principles. To assess the real-world performance of \sg{}-based BCM we plan to integrate the present algorithm into the Parallel Particle-Mesh (PPM) Library \cite{sbalzarini2006ppm,sbalzarini2010abstractions,awile2010toward,awile2013domain} and test its performance in massively parallel real-world simulations where load imbalance is mostly due to the dynamics of the simulated phenomenon.

\newpage
\bibliography{DLB_Bib}
\newpage

\section{Appendix A: Proof of Theorem 1}
\label{sec:proof}
We prove the expected performance of a \sg{}-based DLB algorithm working on indivisible real-weight loads under the conditions listed in section~\ref{sec:bcm}. We need the following lemmata:
\begin{mylemma}
\label{lemma:e_c}
The error $e_c$ in every matching $[u:v]$ is always zero in the \textit{continuous} case.
\end{mylemma}

\textit{Proof.} Let $\xi_u$ and $\xi_v$ be the local load vectors on $u$ and $v$, respectively. The evolution of the load vector is a linear system and can be written as $\xi^{(t)} = \xi^{(t-1)}\mathbf{M}^{(t)}$. Further, the evolution of the loads on node $u$ can be formulated as:
\begin{eqnarray}
	 \xi_{u}^{(t)}  &=&\xi_u^{(t-1)}+ \sum_{v: \{u,v\}\in E} \left (\xi_v^{(t-1)} \mathbf{M}_{v,u}^{(t)} - \xi_u^{(t-1)} \mathbf{M}_{u,v}^{(t)}  \right ) \\
	 		   &= &\xi_u^{(t-1)}+ \sum_{v: \{u,v\}\in \mathbf{M}^{(t)}} \left ( \frac{1}{2} \xi_v^{(t-1)} - \frac{1}{2} \xi_u^{(t-1)} \right ) .
\end{eqnarray}
The evolution of the load vector is a Markov chain and its convergence speed is closely related to its spectral gap $(1-\lambda (\mathbf{M}))$. In the continuous case after a matching $[u:v]$ both $\xi_u$ and $\xi_v$ will be the same. Since $e_c=|\xi_u-\xi_v|=0$, we will always have a perfectly balanced state after each matching. $\blacksquare$

\begin{mylemma}
\label{lemma:d}
Let $e_f$ denote the load imbalance in the indivisible-weight case after balancing local loads $\xi_u$ and $\xi_v$ on nodes $u$ and $v$, respectively. The difference $d$ between $e_f$ and $e_c$ after balancing a matched edge equals to $e_f$.
\end{mylemma}

\textit{Proof.} From Lemma \ref{lemma:e_c} we have $e_c=0$ for every matching, hence $d=|e_f - e_c|= e_f$. $\blacksquare$

\begin{mylemma} 
\label{lemma:proof}
Let the load vector $l:=\{l_1,l_2,\,\ldots,\,l_n\}$ with $l_1\geq l_2 \geq,\,\ldots,\,\geq l_n$. The maximum difference $|d_{max}|:=max(|e_f - e_c|)$ obtained by \sg{} is $\left | d_{max} \right | \leq \frac{l_1}{2}$. 
\end{mylemma}

\textit{Proof.} Consider the worst case where all loads are equal to each other, $l_1=l_2=\ldots=l_l=\mathcal{L}$. In this case, the minimum discrepancy achieved by \sg{} is maximized. This is due to the fact that all loads carry maximum possible weight compared to each other. The algorithm places the first load on processor A, which is chosen arbitrarily. The total weights of processors A and B hence are $\mathcal{L}$ and $0$, respectively, for any B. The ideal load distribution would correspond to $\mathcal{L}/2$ on each processor. Thus, the discrepancy is $\mathcal{L}/2$ and it will remain at most $\mathcal{L}/2$ until all loads are placed. $\blacksquare$ 

\medskip
Now, we prove that the present case and \sg{} fulfill all requirements stated in section \ref{sec:bcm}: \\
\textit{Proof of requirement 1}: By definition, the load weights do not change during an \textit{offline} DLB process. Only their hosts (i.e., nodes) change.  $\blacksquare$ \\
%\textit{Proof of  \ref{sec:fixed-weight}.2}:  The algorithm puts first $\alpha$ loads on node $u$ such that:
%\begin{eqnarray}
%	 \sum_{i=1}^{\alpha}l_i = \tilde{l}_{u,v} - \epsilon_1, \label{eq:1}\\
%	 \sum_{i=1}^{\alpha}l_i +l_{\alpha+1}  = \tilde{l}_{u,v} + \epsilon_2 \label{eq:2},
%\end{eqnarray}
%where $\{\epsilon_1, \epsilon_2 \} \geq 0$. By subtracting (\ref{eq:1}) from (\ref{eq:2}) we get:
%\begin{equation}
%	 l_{\alpha+1}  = \epsilon_2 - \epsilon_1 \label{eq:3}.
%\end{equation}
\textit{Proof of requirement 2}: We consider the algorithms \sg{} and \texttt{Greedy}, that try to balance the loads as evenly as possible. $\blacksquare$ \\
\textit{Proof of requirement 3}: To show that $\mathbb{E}[e_{u,v}^{(t)}]=0$, we can look at the two-bin case between $u$ and $v$. Due to the symmetry $e_{u,v}^{(t)}=-e_{v,u}^{(t)}$, the expected error on an edge is always zero. $\blacksquare$ \\
\textit{Proof of requirement 4}: We closely follow the proof given in Ref.~\cite{sauerwald2012tight}, but we have to adjust the concentration bounds for the error. In Ref.~\cite{sauerwald2012tight}, unit loads are considered, hence $e_{u,v}^{(t)} \in \{-1/2,0,1/2\}$, and errors on different edges are independent of each other. In the present case of indivisible real-valued loads, $e_{u,v}^{(t)}$ is also independent of errors on other edges and, due to Lemma \ref{lemma:proof}, it holds that $\{-\frac{l_{\max}}{2}\leq e_{u,v}^{(s)}\leq \frac{l_{\max}}{2}\}$, where $l_{\max}$ is the largest single load in the entire network. In words, the maximum error on any edge is bounded by the largest load in the network. This enables us to use also Lemma 2.13 from  Ref.~\cite{sauerwald2012tight}: 
\begin{mylemma}[\cite{sauerwald2012tight}, Lemma 2.13]
\label{lemma:final}
Fix an arbitrary load vector $x^{(0)}$. Consider two rounds $t_1\leq t_2$ and assume that the time-interval $[0,t_1]$ is $(K,1/(2n))$-smoothing. Then, for any node $w \in V$ and $\delta > 1/n$, it holds that
 	\begin{equation}
			\Pr\left [ \left | x_w^{(t_1)} - \bar{x} \right | \geq \delta \right ] \leq 2 \cdot \exp \left ( - \left ( \delta - \frac{1}{2n} \right )^{2}/4\right ).
	\end{equation} 

\end{mylemma}

Using Lemmata \ref{lemma:first}, \ref{lemma:proof}, and \ref{lemma:final}, and following the same derivation in Ref.~\cite{sauerwald2012tight} (Lemmata 2.12 and 2.13 therein), it follows that Theorem \ref{theorem:result} also holds for a BCM with indivisible, real-weight loads.  $\blacksquare$ \\

\section{Appendix B: Proof of Theorem 2}

We have defined
\begin{equation}
 U_i^{(k)}:=\sum_{j\in S} W_j \, , 
 \label{eqn:uik}
\end{equation}
where $S$ is the list of the ball weights of size $J$ in $U_i^{(k)}$. If there are statistically enough number of balls, i.e. $J\geq 30$, we can write Eq.~\ref{eqn:uik} as:
\begin{equation}
	U_i^{(k)}:=J\cdot \bar{W}.
\end{equation}
where $\bar{W}$ is the mean of all ball weights $W_{1\ldots n}$. 
%Note that we can write $U_{i+1}^{(k)}$ as $U_i^{(k)}:=(J+1)\cdot \bar{W}$ only if the last ball weight $W_{i+1}$ is \textit{randomly} selected. Else, the mean of the $i+1$ thrown balls $\bar{W_i}$ may not be statistically equal to $\bar{W}$. \sg{} helps us exploit the fact that $\bar{W_i}\neq\bar{W}$. More details on \sg{} can be found in the following sections. \\
Further, the discrepancy after placing $i$ out of $m$ balls is:
\begin{equation}
G_i=\max_{k} U_i^{(k)} - \min_{k} U_i^{(k)}.
\end{equation} 
where $k=\{1,\,\ldots ,\,n\}$.
To make the analysis easier we put the tag ``heaviest" on the heaviest bin $U_{\texttt{heaviest}}$ and a ``switch" happens if after throwing next ball, another bin takes the tag ``heaviest," i.e. another bin becomes the heaviest bin. If no switch occurs, the heaviest bin is still the same but the discrepancy is reduced by the weight of the next ball $W_{i+1}$. Moreover, in the ``switch" case $G_{i}$ can change at most by $W_{i+1}$. We examine the offline weighted-balls-into-bins problem in two different test cases, namely two-bin and $n$-bin case where $n>2$.

\subsection{Two-bin case}
Two-bin case is quite easy to understand and very important in practical distributed dynamic load balancing protocols where a nearest neighbor is chosen and two sides balance their loads. In such scenarios, the solution of the two-bin problem is analogous to the local dynamic load balancing solution.

We start our analysis after putting the first random ball in either of bins $U^{(1)}$ or $U^{(2)}$. Later, we assume that $U^{(1)}$ is heavier than $U^{(2)}$ after throwing $i^{\text{th}}$ ball. The discrepancy is defined as $G_i=U_i^{\texttt{heaviest}} - U_i^{(2)}$. Later on depending on the following random ball $W_{i+1}$ a ``switch" may or may not occur.  We can define $G_{i+1}$ in two cases as follows:
\begin{itemize}
	\item ``no switch": $G_{i+1}=U_{i+1}^{\texttt{heaviest}} - U_{i+1}^{(2)}$.
	\item ``switch": $G_{i+1}=U_{i+1}^{\texttt{heaviest}} - U_{i+1}^{(1)}$.
\end{itemize} 

We are interested in the difference between the consecutive discrepancies $\Delta G_{i+1} = G_{i}-G_{i+1}$ in both cases. \\
1) \textit{``No-switch" case}: We put the next ball $W_{i+1}$ in $U^{(2)}$. Thus, the total weight of $U^{(1)}$ does not change but the discrepancy is reduced by $W_{i+1}$:
\begin{align}
	\Delta G_{i+1}&=U_{i}^{(1)}- U_{i}^{(2)} - (U_{i+1}^{(1)} - U_{i+1}^{(2)}) \\ \nonumber
	&=U_{i+1}^{(2)}- U_{i}^{(2)} \\ 
	&= W_{i+1}.
	\label{eqn:twobin_ns}
\end{align}
2) \textit{``Switch" case}: Let us assume that $U_{i}^{(1)}$ has $J$ balls in it, whereas $U_{i}^{(2)}$ contains $K$ balls such that $J+K=i$. We put the next ball $W_{i+1}$ in $U^{(2)}$ and the tag ``heaviest" switches from $U^{(1)}$ to $U^{(2)}$. Now, after $i+1$ balls $U^{(2)}$ contains $K+1$ balls. The total weight of $U^{(1)}$ does not change again, $U_{i}^{(1)}=U_{i+1}^{(1)}$ and the discrepancy difference is upper bounded by $W_{i+1}$ only if  $U_{i}^{(1)}=U_{i}^{(2)}$. The discrepancy difference is as follows:
\begin{align}
	\Delta G_{i+1}&= U_{i}^{(1)}- U_{i}^{(2)} - \left | U_{i+1}^{(1)} - U_{i+1}^{(2)} \right | \\ \nonumber
	&=U_{i}^{(1)}- U_{i}^{(2)} + U_{i+1}^{(1)} - U_{i+1}^{(2)} \\\  \nonumber
	&=2 \cdot U_{i}^{(1)} - U_{i}^{(2)} - U_{i+1}^{(2)} \\ \nonumber
	&= 2\cdot U_{i}^{(1)} - 2\cdot U_{i}^{(2)} - W_{i+1}   \\
	&\leq W_{i+1}.
	\label{eqn:twobin_switch}
\end{align}

 If $i$ is large enough to do statistical analysis, we can re-write equation~\ref{eqn:twobin_switch} as follows:
\begin{align}
	\nonumber
	\Delta G_{i+1}&=2\cdot U_{i}^{(1)} - U_{i}^{(2)} - U_{i+1}^{(2)} \\ \nonumber
	&\simeq 2J \cdot \bar{W} - K \cdot \bar{W} - (K+1) \cdot \bar{W} \\
	&\simeq (2J-2K+1)\cdot \bar{W} .
	\label{eqn:twobin_switch2}
\end{align}

If $\mathcal{D}$ is uniformly random and $m$ is large enough and even, $J=K$ holds. On the other hand, for odd $m$, $J\simeq K$. Thus, we can combine equations~\ref{eqn:twobin_switch} and \ref{eqn:twobin_switch2} into:
\begin{align}
	\Delta G\simeq \bar{W} \leq W_{i+1}.
	\label{eqn:twobin_switch3}
\end{align}

For other distributions the relation between $J$ and $K$ depends on the standard deviation of $\mathcal{D}$. $\blacksquare$

\subsection{$n$-bin case}
The extension of two-bin problem to $n$-bin problem is straightforward. We add an additional tag ``lightest." In the two-bin case, the bin with the ``lightest" tag is trivial and the tag is not used. Yet, here we take advantage of having this second tag. One important fact to consider is the existence of other \textit{intermediate} bins whose total weights lie between the heaviest and lightest bin. The ``switch" and ``no-switch" of the ``heaviest" tag can be written as follows:\\
1) \textit{``No-switch" case}: We put the next ball $W_{i+1}$ into the lightest bin $U_{i}^{\texttt{lightest}}$. Since we have $n$ bins, an intermediate bin might become the lightest bin after $i+1$ balls or $U_{i}^{\texttt{lightest}}$ is increased by $W_{i+1}$. Nevertheless, regardless of the value of $W_{i+1}$ it holds that $U_{i+1}^{\texttt{lightest}}>U_{i}^{\texttt{lightest}}$. On the other hand, $U_{i}^{\texttt{heaviest}} = U_{i+1}^{\texttt{heaviest}}$ since a ``switch" does not occur. Thus, the discrepancy difference is written as follows:
\begin{align}
\nonumber
	\Delta G_{i+1}&= U_{i}^{\texttt{heaviest}} - U_{i}^{\texttt{lightest}} - (U_{i+1}^{\texttt{heaviest}} - U_{i+1}^{\texttt{lightest}})  \\ \nonumber
	&= U_{i+1}^{\texttt{lightest}} - U_{i}^{\texttt{lightest}} \\
	&\leq W_{i+1}.
	\label{eqn:mbin_ns}
\end{align}

since the maximum $\Delta G$ is achieved only if the previous lightest bin gets $W_{i+1}$ and is still the lightest. 
For large enough $i$, we can reformulate equation~\ref{eqn:mbin_ns} by introducing a statistical upper bound on the discrepancy difference $\Delta G_{i+1}$. \\
\begin{align}
	\nonumber
	\Delta G_{i+1} &= U_{i+1}^{\texttt{lightest}} - U_{i}^{\texttt{lightest}}   \\  \nonumber
	&\simeq (K+1) \cdot \bar{W} - K \cdot \bar{W} \\
	&\simeq \bar{W}	.
	\label{eqn:mbin_ns_stats}
\end{align}

where $K<i$ is the number of balls in $U_{i}^{\texttt{lightest}}$. We do a similar combination as in the two-bin case and obtain using equations~\ref{eqn:mbin_ns} and \ref{eqn:mbin_ns_stats}: \\
\begin{align}
	\Delta G_{i+1}\simeq \bar{W} \leq W_{i+1}.
	\label{eqn:mbin_switch3}
\end{align}
2) \textit{``Switch" case}: Switching the ``heavier" tag to another bin states that 
\begin{align}
	\nonumber 
	U_{i+1}^{\texttt{heaviest}}>U_{i}^{\texttt{heaviest}} ,
\end{align}

and 
\begin{align}
	U_{i+1}^{\texttt{heaviest}}=U_{i}^{\texttt{lightest}}+W_{i+1}. 
	\label{mbin_switch_ui1}
\end{align}
Moreover, a previously intermediate bin becomes the lightest bin:  
\begin{align}
	\nonumber
	U_{i+1}^{\texttt{lightest}} \geq U_{i}^{\texttt{lightest}}.
\end{align} 

Yet, the relation is now
\begin{align}
	\nonumber
	U_{i+1}^{\texttt{lightest}}\neq U_{i}^{\texttt{lightest}}+W_{i+1}.
\end{align} 

Hence, the discrepancy difference is: 
\begin{align}
\nonumber
	\Delta G_{i+1}&=  U_{i}^{\texttt{heaviest}} - U_{i}^{\texttt{lightest}} - ( U_{i+1}^{\texttt{heaviest}} - U_{i+1}^{\texttt{lightest}} )  \\
	&= U_{i}^{\texttt{heaviest}} - U_{i+1}^{\texttt{heaviest}}  + U_{i+1}^{\texttt{lightest}} - U_{i}^{\texttt{lightest}} .
	\label{eqn:mbin_switch}
\end{align}

We cannot further reduce the equation~\ref{eqn:mbin_switch} since each term therein depends on the specific weight sampling scored from $\mathcal{D}$. However, we can tightly bound the maximum $\Delta G_{i+1}$ by considering all intermediate bins having the same total weight as $U_{i}^{\texttt{lightest}}$ after $i^{\text{th}}$ ball. This way, we imply:
\begin{align}
	U_{i+1}^{\texttt{lightest}}&=U_{i}^{\texttt{lightest}} , \label{mbin_switch_ui2} \\	
	W_{i+1}&\geq U_{i}^{\texttt{heaviest}} - U_{i}^{\texttt{lightest}} .
	\label{mbin_switch_wi1}
\end{align} 

Thus, substituting equations~\ref{mbin_switch_ui1}, \ref{mbin_switch_ui2} and \ref{mbin_switch_wi1} in equation~\ref{eqn:mbin_switch}, $\Delta G$ is upper bounded as follows:
\begin{align}
\nonumber
	\Delta G_{i+1}&= U_{i}^{\texttt{heaviest}} - U_{i+1}^{\texttt{heaviest}}  + U_{i+1}^{\texttt{lightest}} - U_{i}^{\texttt{lightest}}  \\ \nonumber
	&= U_{i}^{\texttt{heaviest}} - U_{i+1}^{\texttt{lightest}} \\
	&\leq W_{i+1}.
	\label{eqn:mbin_switch_bound}
\end{align}

A statistical investigation of the upper bound on $\Delta G_{i+1}$ makes a randomly selected $W_{i+1}\rightarrow \bar{W}$. Thus,
\begin{align}
	\Delta G_{i+1} \simeq \bar{W}\leq W_{i+1}. \qquad \blacksquare
	\label{eqn:mbin_switch_tightbound}
\end{align} 

\subsection{Lower bound on $G_m$}
 Deriving an upper bound for all possible $G_m$ is hard. However, we can derive a non-trivial lower bound for both cases; namely, when each thrown ball triggered a ``switch" and where none of the balls caused a ``switch": \\

$\left . \begin{matrix}
 & \Delta G_2 = G_1 - G_2  \leq W_2 & \\
 & \Delta G_3 = G_2 - G_3  \leq W_3 & \\
 & ... & \\ 
 & \Delta G_m = G_{m-1} - G_m  \leq W_n & 
\end{matrix}\right \}$ 
We add all $\Delta G_i$ values. \\

This gives us 

\begin{align}
\nonumber
 G_1 - G_m \leq \sum_{i=2}^{m} W_i,
 \end{align}
 
 where $G_1=|W_1|$ and thus
  \begin{align}
   G_m \geq W_1 - \sum_{i=2}^{m} W_i \geq 0. 
   \label{eqn:final_gn}
 \end{align}
 
For statistically large $m$, we can rewrite Eq.~\ref{eqn:final_gn} as
 \begin{align}
   G_m \geq 2W_1 - \bar{W}m \geq 0.
   \label{eqn:final_gn_stat}
 \end{align}

\section{Appendix C: Benchmarking the \sg{} Algorithm}

We implement both \gr{} and \sg{} in MATLAB (R2012a, The Mathworks, Inc., Natick, MA, USA). \sg{} uses MATLAB's intrinsic quicksort function to sort the balls according to their weights. The balls are assigned random weights sampled from a uniform distribution over the interval $[0,1]$. Each simulation is repeated 1000 times with different random weights, and we report the mean and standard deviation $\sigma$ of the discrepancy for different numbers of balls and bins. 

\subsection{Increasing $m$}

\begin{figure}
\centering
	\begin{subfigure}[b]{0.49\textwidth}	
	      	\includegraphics[width=\textwidth]{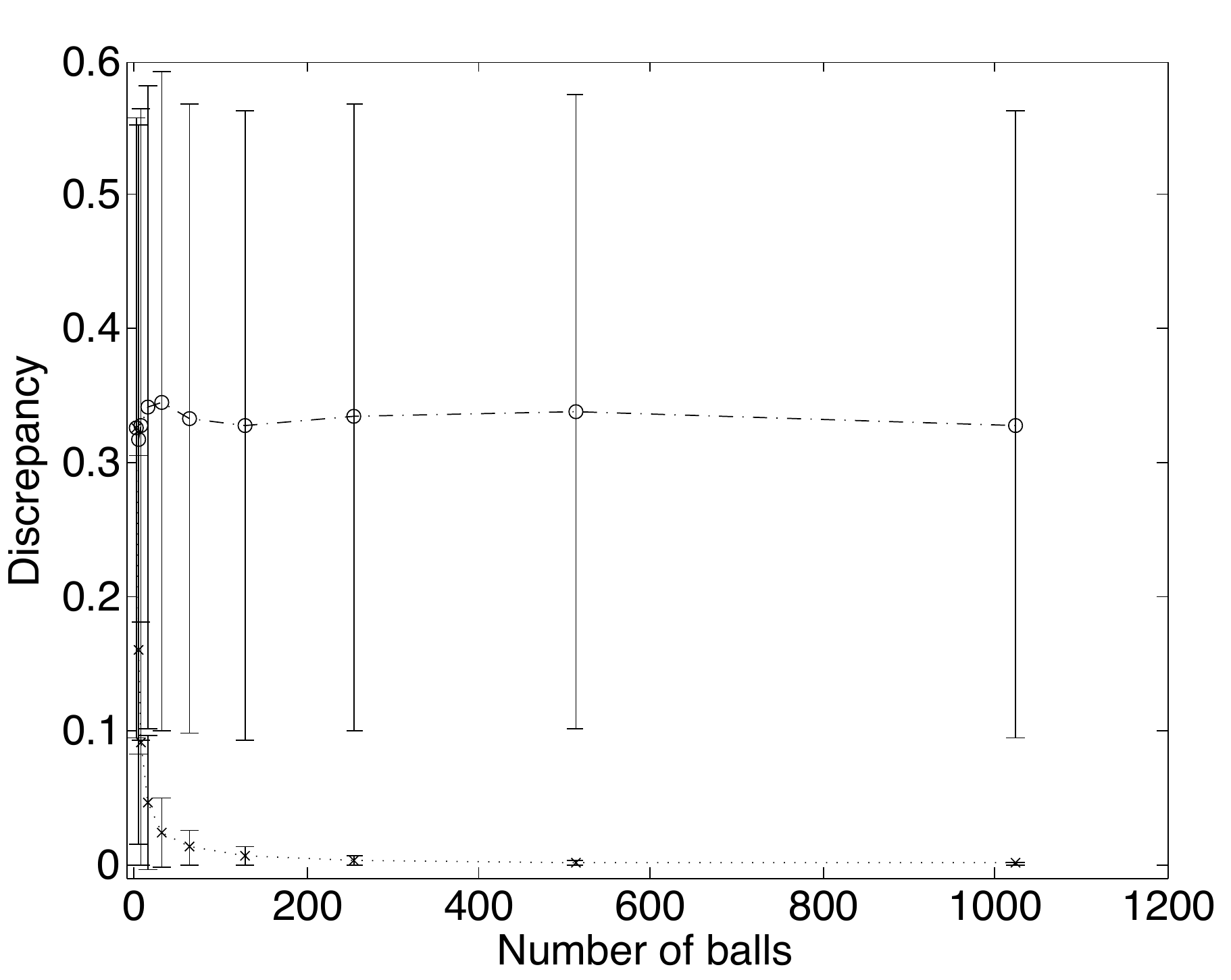} \\
		\centering (a) $n=2$
	\end{subfigure} 
	\begin{subfigure}[b]{0.49\textwidth}	
	      	\includegraphics[width=\textwidth]{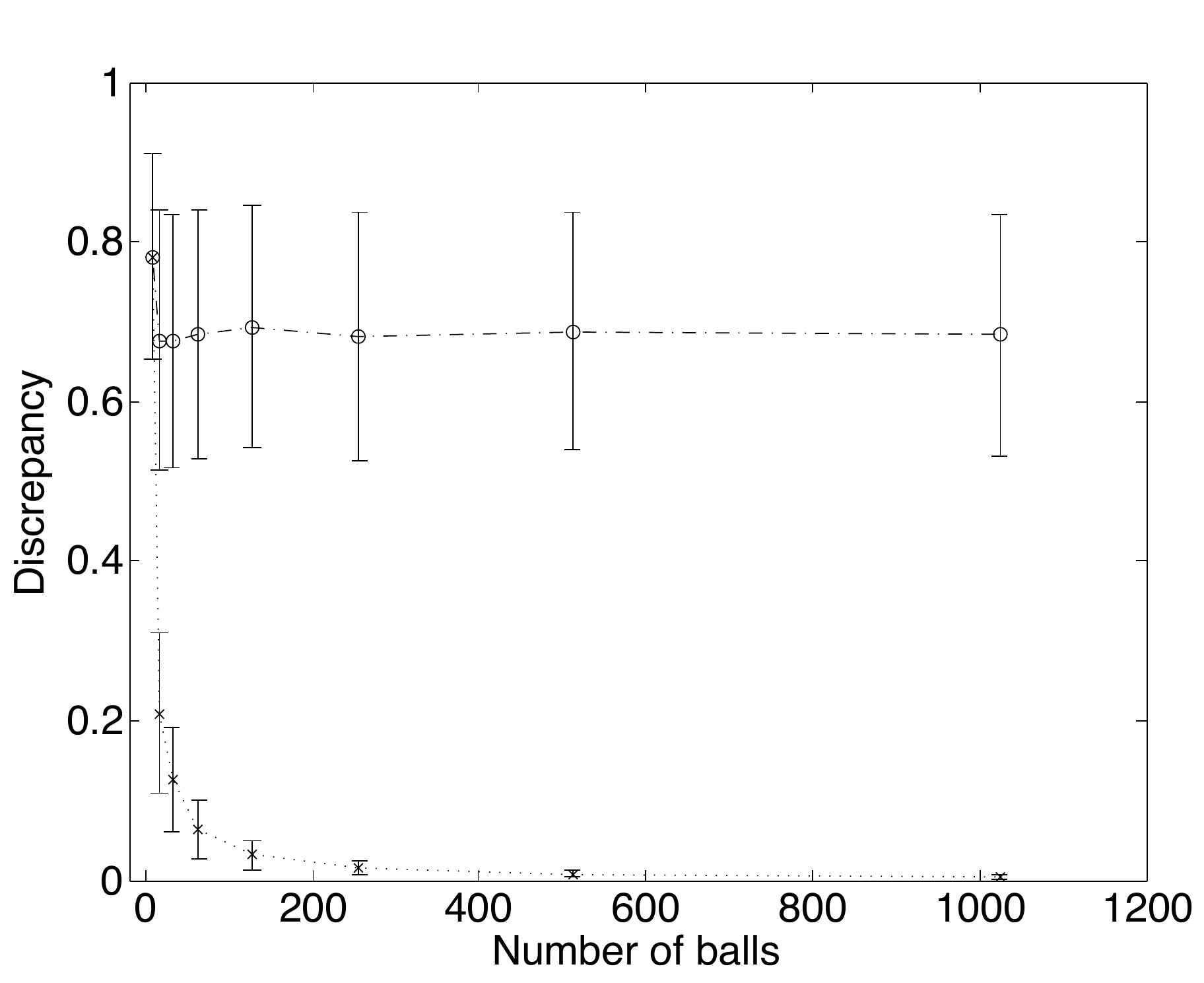} 
		\centering (b) $n=8$
	\end{subfigure} 
	\caption{The discrepancy is shown for each $m$.  On average, the discrepancies achieved by \sg{} (x) are an order of magnitude smaller than those obtained by \gr{} ($\circ$). (a) The case for $n=2$ bins. For $m\geq32$, the average discrepancy ratio between the two algorithms increases to 60. (b) The case for $n=8$ bins. Here, the discrepancy ratio is about 73 for $m\ge 512$.}
	\label{fig:gaps}
\end{figure}

Figure~\ref{fig:gaps} shows the results for $n=\{2,8\}$ bins and varying numbers of balls. The $\sigma$ bars for \gr{} are independent of $m$ with $\sigma=0.23$ for $n=2$ and $\sigma=0.15$ for $n=8$. For \sg{}, the average $\sigma$ is 0.01 for $n=2$ and 0.03 for $n=8$. 

As seen in Fig.~\ref{fig:gaps}, \sg{} outperforms \gr{} in all tested cases, including those with odd numbers of balls. The discrepancy resulting from \sg{} decreases exponentially as the number of balls increases, and it is at least 10 times smaller than the discrepancies obtained by \gr{} when $m\gg n$. For each $n$-bin problem, the standard deviation across the random repetitions of the \gr{} algorithm remains constant. Also, the discrepancy resulting from \gr{} remains almost constant with $m$. 

\subsection{Increasing $n$}

\begin{figure}
\centering
	\begin{subfigure}[b]{0.49\textwidth}	
	      	\includegraphics[width=\textwidth]{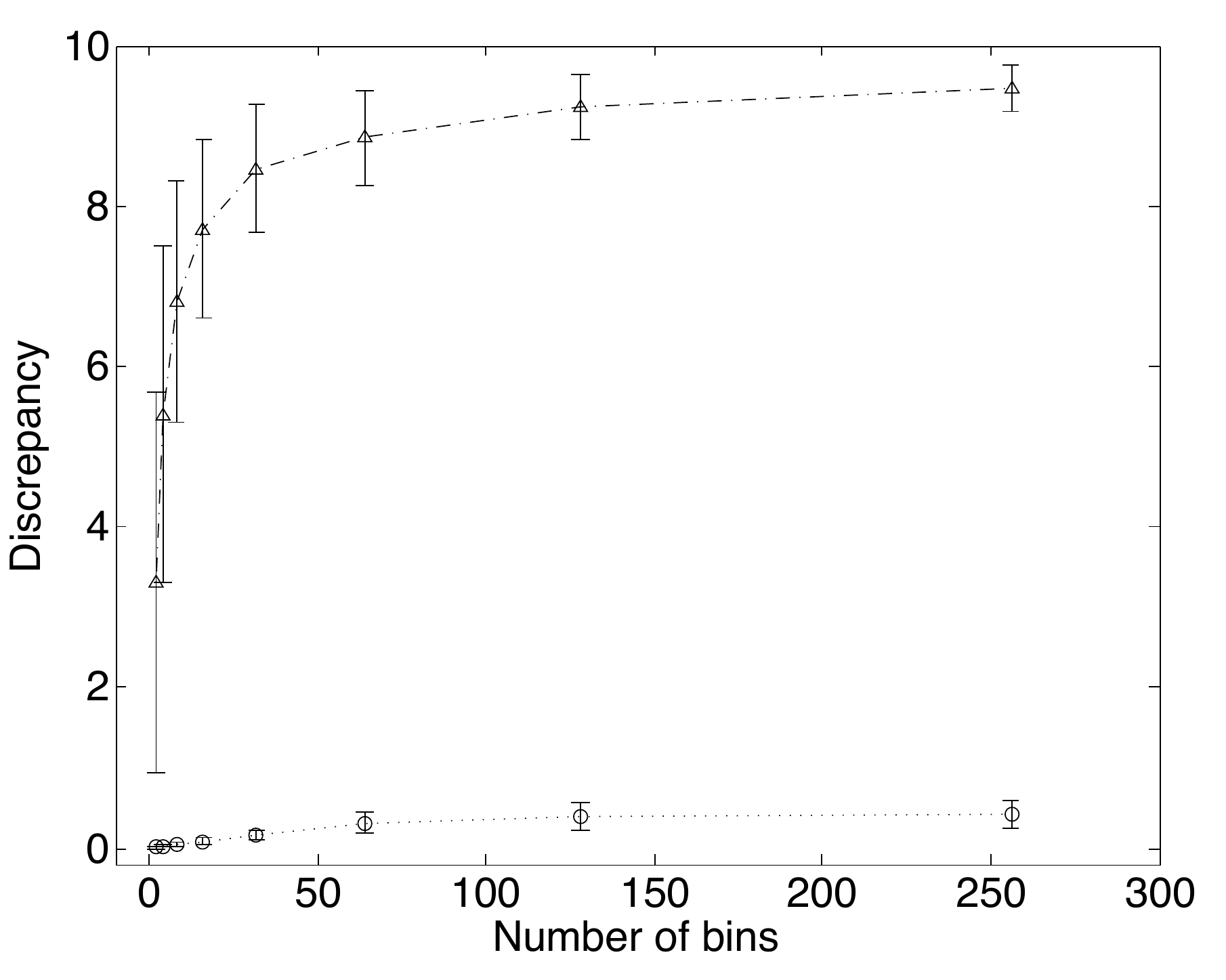} \\
		\centering (a) $m=1024$
	\end{subfigure} 
	\begin{subfigure}[b]{0.49\textwidth}	
	      	\includegraphics[width=\textwidth]{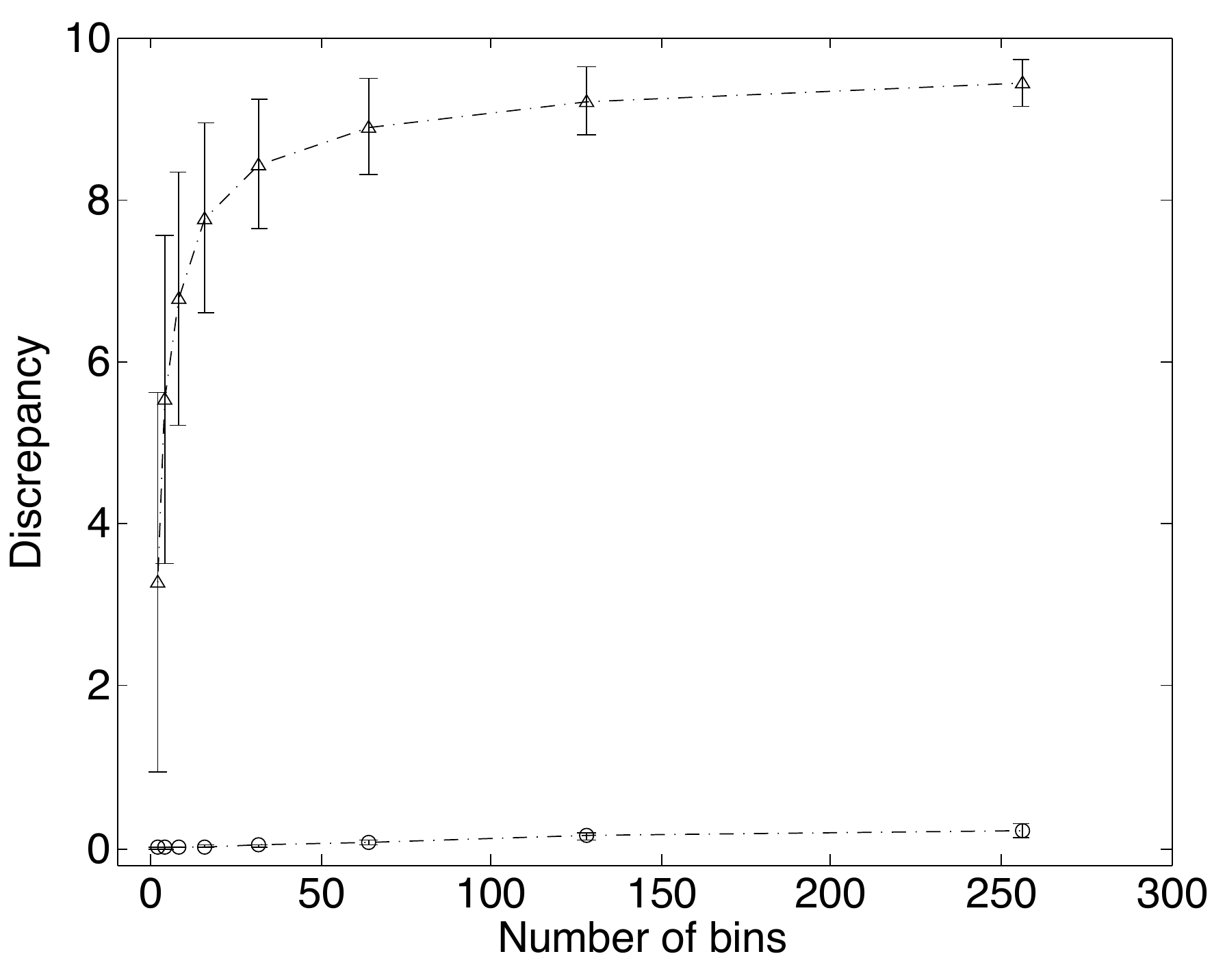} \\
		\centering (b) $m=3027$
	\end{subfigure} 
	\caption{The discrepancy achieved for different numbers of bins and a constant number of balls: (a) 1024 balls, (b) 3027 balls. Results are shown for \sg{} algorithm ($\circ$) and \gr{} ($\triangle$).} 
	\label{fig:gaps_bins}
\end{figure}

 In Fig.~\ref{fig:gaps_bins} we show the dependence of the discrepancy on the number of bins $n$ for $m=\{1024,3027\}$. The discrepancy obtained by \gr{} first increases rapidly and then seems to saturate. That from \sg{} initially increases much slower. This is in line with previous findings~\cite{talwar2007balanced}. Indeed, Talwar \textit{et al.}~\cite{talwar2007balanced} show that the discrepancy depends on both the distribution from which the weights are sampled, and on $n$.

\subsection{Timings}

We perform runtime measurements for the two-bin problem with $m=2^{13}$. The experiment is repeated 100 times and averages are recorded. All test runs are conducted on a Macbook Pro (MacOS X 10.7.5) with a quad-core 2.3\,GHz Intel Core i7 processor and 8\,GB 1600\,Mhz DDR3 memory. Both algorithms require approximately the same time to solve the two-bin problem. For placing $2^{13}$ balls, 0.1950\,s are needed by \sg{} and 0.1948\,s by \gr{}. Thus, sorting adds an overhead of about 2\,ms, which is $0.02\%$ of the total runtime. Increasing $n$ has no substantial effect on the final runtime as long as $m\gg n$. \\

\end{document}